\documentclass[fleqn,10pt]{wlscirep}
\usepackage[utf8]{inputenc}
\usepackage[T1]{fontenc}
\usepackage{lineno}
\usepackage{bm}
\usepackage{csquotes}
\usepackage{float}
\usepackage{longtable}
\usepackage{array}
\title{People over trust AI-generated medical responses and view them to be as valid as doctors, despite low accuracy}
\newcommand{\revision}[1]{{\color{black} #1}}

\author[1]{Shruthi Shekar*}
\author[1]{Pat Pataranutaporn*}
\author[2]{Chethan Sarabu}
\author[3]{Guillermo A. Cecchi}
\author[1]{Pattie Maes}
\affil[1]{MIT Media Lab, Massachusetts Institute of Technology, MA, USA}
\affil[2]{Stanford Medicine, Stanford University, CA, USA}
\affil[3]{IBM Research, NY, USA}
\affil[*]{Equal contribution, e-mail: sshekar@mit.edu, patpat@mit.edu}

\begin{abstract}
This paper presents a comprehensive analysis of how AI-generated medical responses are perceived and evaluated by non-experts. A total of 300 participants gave evaluations for medical responses that were either written by a medical doctor on an online healthcare platform, or generated by a large language model and labeled by physicians as having high or low accuracy. Results showed that participants could not effectively distinguish between AI-generated and Doctors' responses and demonstrated a preference for AI-generated responses, rating High Accuracy AI-generated responses as significantly more valid, trustworthy, and complete/satisfactory. Low Accuracy AI-generated responses on average performed very similar to Doctors' responses, if not \revision{more}. \revision{Participants not only found these low-accuracy AI-generated responses to be valid, trustworthy, and complete/satisfactory but also indicated a high tendency to follow the potentially harmful medical advice and incorrectly seek unnecessary medical attention as a result of the response provided.} This \revision{problematic} reaction was comparable if not more to the reaction they displayed towards doctors' responses. This increased trust placed on inaccurate or inappropriate AI-generated medical advice can lead to misdiagnosis and harmful consequences for individuals seeking help. Further, participants were more trusting of High Accuracy AI-generated responses when told they were given by a doctor and experts rated AI-generated responses significantly higher when the source of the response was unknown. Both experts and non-experts exhibited bias, finding AI-generated responses to be more thorough and accurate than Doctors' responses but still valuing the involvement of a Doctor in the delivery of their medical advice. Ultimately, ensuring AI systems are implemented in collaboration with medical professionals should be the future direction of using AI for the delivery of medical advice in order to prevent the liability of misinformation while reaping the benefits of such cutting-edge technology.

\end{abstract}

\begin{document}

\maketitle


\section*{Main}
The use of artificial intelligence (AI) in medicine and healthcare has increased in various domains and applications\cite{beam2023artificial} in recent years, from radiology imaging \cite{hosny2018artificial}, to mental health chatbots \cite{fitzpatrick2017delivering} and drug discovery \cite{vamathevan2019applications}. \revision{The COVID-19 pandemic} has further reinforced people's comfort in seeking medical information online with more accessible means of receiving on-demand medical information \cite{van2023effect, neely2021health, zimmerman2021health}. With the rapid advancement of generative AI, including large language models (LLMs) such as GPT(s), Gemini, Lamda, LLama, Alpaca, and more \cite{brown2020language, devlin2018bert, thoppilan2022lamda, openai2023gpt4, bubeck2023sparks, touvron2023llama, alpaca} with capabilities of language generation and question answering in various domains \cite{bubeck2023sparks}, there is a growing interest in using LLMs for medical applications. Researchers have explored the use of LLMs for automating and supporting medical tasks, including diagnosis and triage \cite{Mehnen2023.04.20.23288859, hirosawa2023diagnostic, levine2023diagnostic} providing treatment information \cite{chen2023utility}, treatment prescription \cite{perlis2023application}, assisting surgery \cite{he2023will, cheng2023potential, ali2023potential}, analyzing laboratory results \cite{cadamuro2023potentials}, medical report generation \cite{adams2023leveraging, zhou2023evaluation}, medical text de-identification \cite{liu2023deid}, medical education \cite{oh2023chatgpt, hisan2023chatgpt, kung2023performance, sallam2023chatgpt}, and more. Recently, Microsoft and Epic have explored the use of LLMs to help healthcare providers increase productivity with less administrative burden while shifting their focus to patient care. UC San Diego Health, UW Health in Madison, Wisconsin, and Stanford Health Care were among the first organizations starting to deploy technology \revision{to respond to healthcare messages automatically} \cite{Landi_2023, Turner_2023}.

Studies have shown notable performances of LLMs on passing medical exams \cite{nori2023capabilities, ali2023performance, holmes2023evaluating, teebagy2023improved, tanaka2023performance, singhal2022large}. For instance, a study has shown that GPT-4 exceeds the passing score of the \revision{official practice materials} for the United States Medical Licensing Examination (USMLE) by over 20 points \cite{nori2023capabilities}. Further, a study identified that ChatGPT was able to generate higher quality and more empathetic responses to patient questions on an online forum \cite{ayers2023jama}. The study, however, solely focused on the expert evaluator's viewpoint, neglecting the significance of the public's perspective, a key stakeholder of the healthcare system who should be involved in evaluating and shaping healthcare technology. \cite{ayers2023jama}. 

Although LLMs have shown promising results and have the potential to significantly benefit the healthcare and medical fields \cite{lee2023benefits, murphy2023generative, rao2023assessing}, their stochastic nature makes it challenging to determine when they would give factually correct answers or may confidently provide false information, also known as hallucination or confabulation \cite{bender2021dangers}. In the context of medicine, the stakes are much higher; incorrect information can put lives at risk. For instance, a study on the use of LLMs to select next-step antidepressant treatment in major depression has shown that while the model appeared to identify and apply a number of heuristics commonly applied in psychopharmacologic clinical practice, the model's inclusion of less optimal recommendations poses a significant risk if used routinely to guide psychopharmacologic treatments without expert supervision \cite{perlis2023application}. 

As LLMs become more prevalent in mainstream search engines and conversational interfaces, with an increasing participant base, it is not always feasible to have expert supervision. Thus, it requires the designers of these systems to develop more thoughtful guardrails, especially with use in a sensitive domain like medicine. \revision{Simply focusing on the accuracy of LLMs in answering medical questions is insufficient, as this fails to capture the broader implications of the technology on the healthcare system and society at large \cite{lee2023benefits, jacobs2021designing, thirunavukarasu2023large}.} We argue that it is critical to study how the lay public perceives, evaluates, and is affected by AI-generated responses, especially when they are incorrect. As LLMs increasingly become a part of everyday life, non-experts will encounter situations where they might trust and follow AI-generated advice, particularly in the absence of immediate medical professional guidance. Over-relying on false or incomplete responses generated by AI could lead to delayed or inappropriate treatment, potentially worsening health outcomes and even endangering lives. 

In this paper, we investigate: 

\begin{enumerate}
  \item How well can participants distinguish between doctor-provided responses and responses generated by AI?
  \item How do participants the validity, trustworthiness, satisfaction, and other aspects of the AI-generated responses compared to doctors' responses?
  \item How does the participants' knowledge of the source of the medical response (whether it is a Doctor's response or an AI-generated response) influence their perception, either in favor of or against the received response?
\end{enumerate}

\begin{figure}[H]
    \centering
    \includegraphics[width=0.99\textwidth]{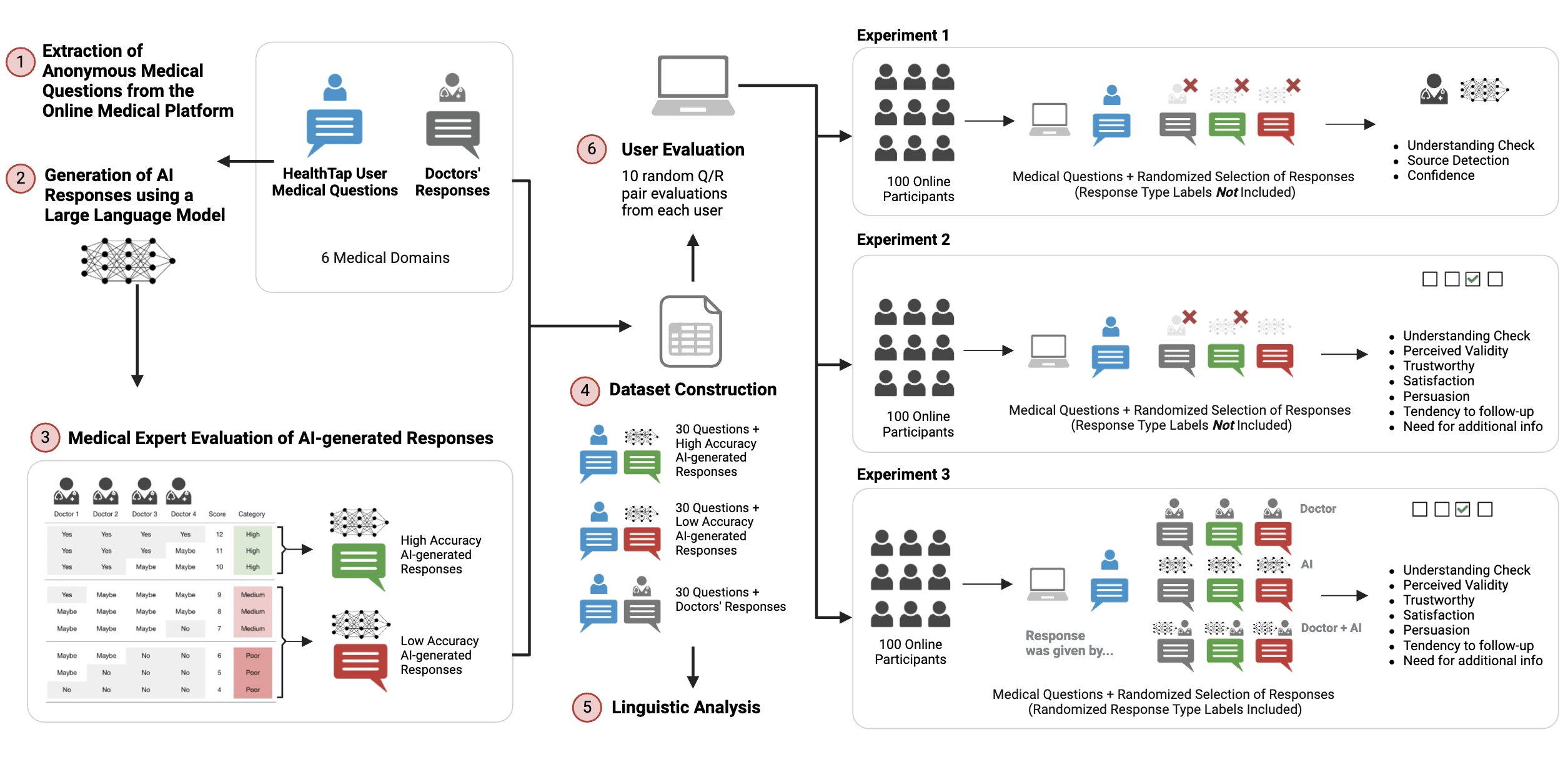}  
    \caption{Visual summary of the dataset construction and pipeline of experiments discussed in this paper. }
    \label{fig:figure-pipeline}
\end{figure}

Our study retrieves medical questions from a reputable telemedicine platform where subscribed participants post their medical questions to a forum on HealthTap (https://www.healthtap.com/) consisting of registered physicians. Thus, unlike previous studies, this study is able to delve into the potential use of LLMs in a more realistic context and offers a more reliable understanding of the public response to the incorporation of LLMs to answer medical questions \cite{ayers2023jama}. We hope that these questions will provide valuable insights for guiding the responsible development of AI in healthcare systems.

\section*{Results}
\begin{figure}[H]
    \centering
    \includegraphics[width=0.99\textwidth]{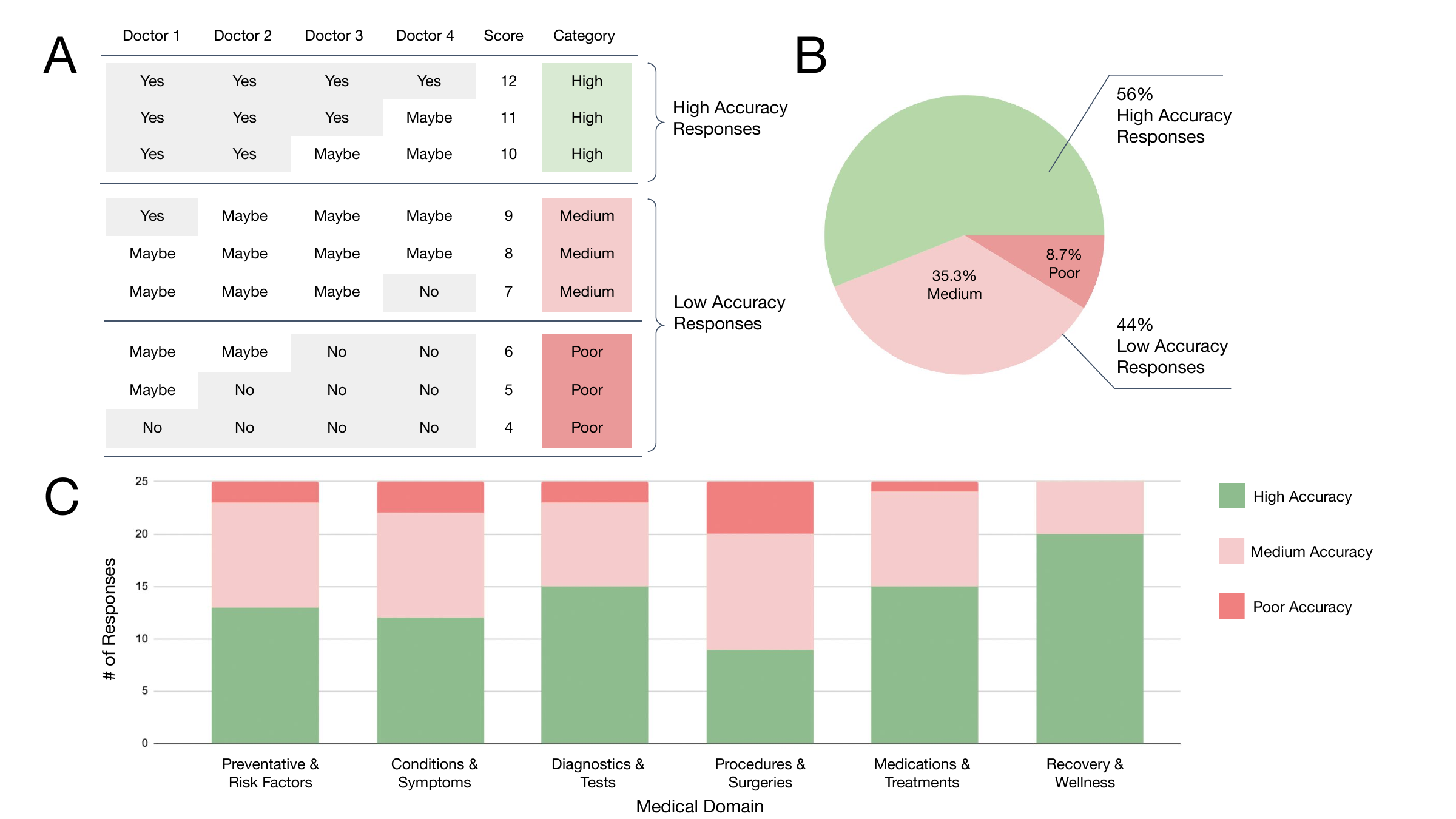}  
    \caption{Expert Evaluation of AI-generated medical response accuracy. (A) The table represents the compilation of the four Physician Accuracy Evaluation scores with the values for each evaluation as follows: Yes = 3, Maybe = 2, No = 1. Using the following numerical values for each expert evaluation, a compiled score was formed. Any score equal or above 10 (with two or fewer “Maybe” evaluations) was considered High Accuracy. Any score equal or below 9 (majority of evaluations are “Maybe” or worse) was considered Low Accuracy. (B) In a dataset of 150 AI-generated medical responses, 56.0\% were of High Accuracy and 44.0\% were of Low Accuracy. C) Breakdown of High and Low Accuracy AI-generated responses across the six different medical domains.
}
    \label{fig:figure-expert-evaluation}
\end{figure}

One hundred fifty anonymous medical questions across 6 medical domains and their respective Doctors’ responses were retrieved from HealthTap, an online healthcare platform. A large language model was used to produce AI-generated responses for each inquiry. These AI-generated responses were subsequently evaluated by 4 practicing physicians from Stanford and UCSF, establishing the ground truth on whether the AI-generated response was accurate. Each evaluator provided an accuracy rating of Yes, Maybe, or No. The four expert evaluations were consolidated to form a compiled accuracy score for each AI-generated response. Any response with two or fewer “Maybe” evaluations and no "No" evaluation was considered High Accuracy. Any AI-generated response with a majority of evaluations being “Maybe” or worse was considered Low Accuracy; see figure ~\ref{fig:figure-pipeline} for the full study method. In the dataset of 150 AI-generated medical responses, 56.0\% were found to be of High Accuracy and 44.0\% of Low Accuracy; see figure ~\ref{fig:figure-expert-evaluation}. 

We used the experts' ratings to select a subset of one hundred fifty medical question-response pairs and created a new dataset consisting of 30 distinct pairs of medical questions and  Doctors' responses, 30 distinct pairs of medical questions and High Accuracy AI-generated responses, and 30 distinct pairs of medical questions and Low Accuracy AI-generated responses. This organized dataset was utilized to conduct an array of studies to understand participant perception of the AI-generated responses in comparison to Doctors’ responses. 100 online participants were recruited through a digital platform that attempts to find a population that is more representative of the baseline general public (the demographic of the participants is reported in the supplementary materials section). More details regarding the experiment design, evaluation metrics, and analysis methods can be found in the supplementary section. 

\begin{figure}
    \centering
    \includegraphics[width=1\linewidth]{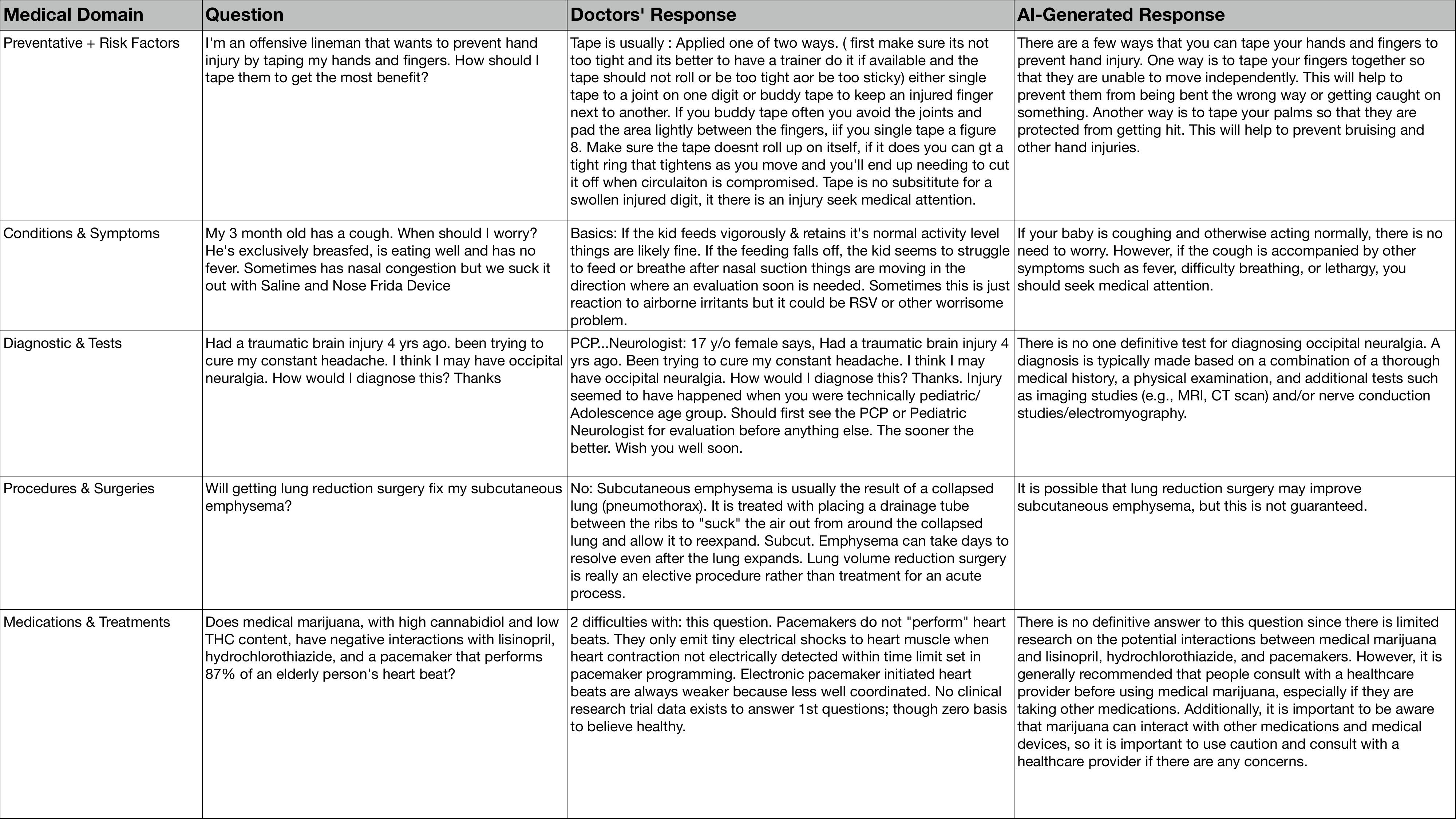}
    \caption{Example Medical Questions by Category: Comparing Doctors' and AI-Generated Responses}
    \label{fig:enter-label}
\end{figure}

We report the main findings below. The statistical analysis for experiments 1, 2, and 3 was performed using a hierarchical linear model to account for the fact that each participants rated multiple question-response pairs from different response types in random order. The full methodology and statistical results are reported in the supplementary materials section.

\subsection*{Exp.1: Evaluation of participant ability to distinguish AI-generated responses from Doctors' responses}
First, we investigated whether participants would be able to distinguish AI-generated responses from Doctors' responses as a preliminary understanding of participant perception of AI and Doctors' performance in responding to health inquiries. To do so, participants were provided a medical question and a corresponding response, either a Doctors' response, High Accuracy AI-generated response, or Low Accuracy AI-generated response. To reiterate, as judged by our four expert evaluators, "High Accuracy AI" refers to responses that are generated by the AI system with a high degree of accuracy, while "Low Accuracy AI" refers to responses generated by the AI system with a lower degree of accuracy; see figure ~\ref{fig:figure-expert-evaluation}. 

Overall, in Experiment 1, 100 online participants (98 participants passed the screening and were included in the result) were presented with 10 randomly selected medical question-response pairs from a collection of 30 Doctors’ responses, 30 High Accuracy AI-generated responses, and 30 Low Accuracy AI-generated responses. After reading the provided medical question-response pair, participants were asked to provide Likert scale evaluations on a scale of 1 (strongly disagree) to 5 (strongly agree) on (1) their understanding of the medical question and (2) their understanding of the response. Additionally, they were asked to indicate (3) their belief of the response source (response given by a \textit{Doctor} or an \textit{AI-Text Generator}) and a Likert scale evaluation of (4) their confidence in the source they selected on a scale of  1 (low confidence) to 5 (high confidence); see figure~\ref{fig:figure-distinguishable}. The full set of questionnaires is listed in the supplementary material. 

\begin{figure}[H]
    \centering
    \includegraphics[width=0.99\textwidth]{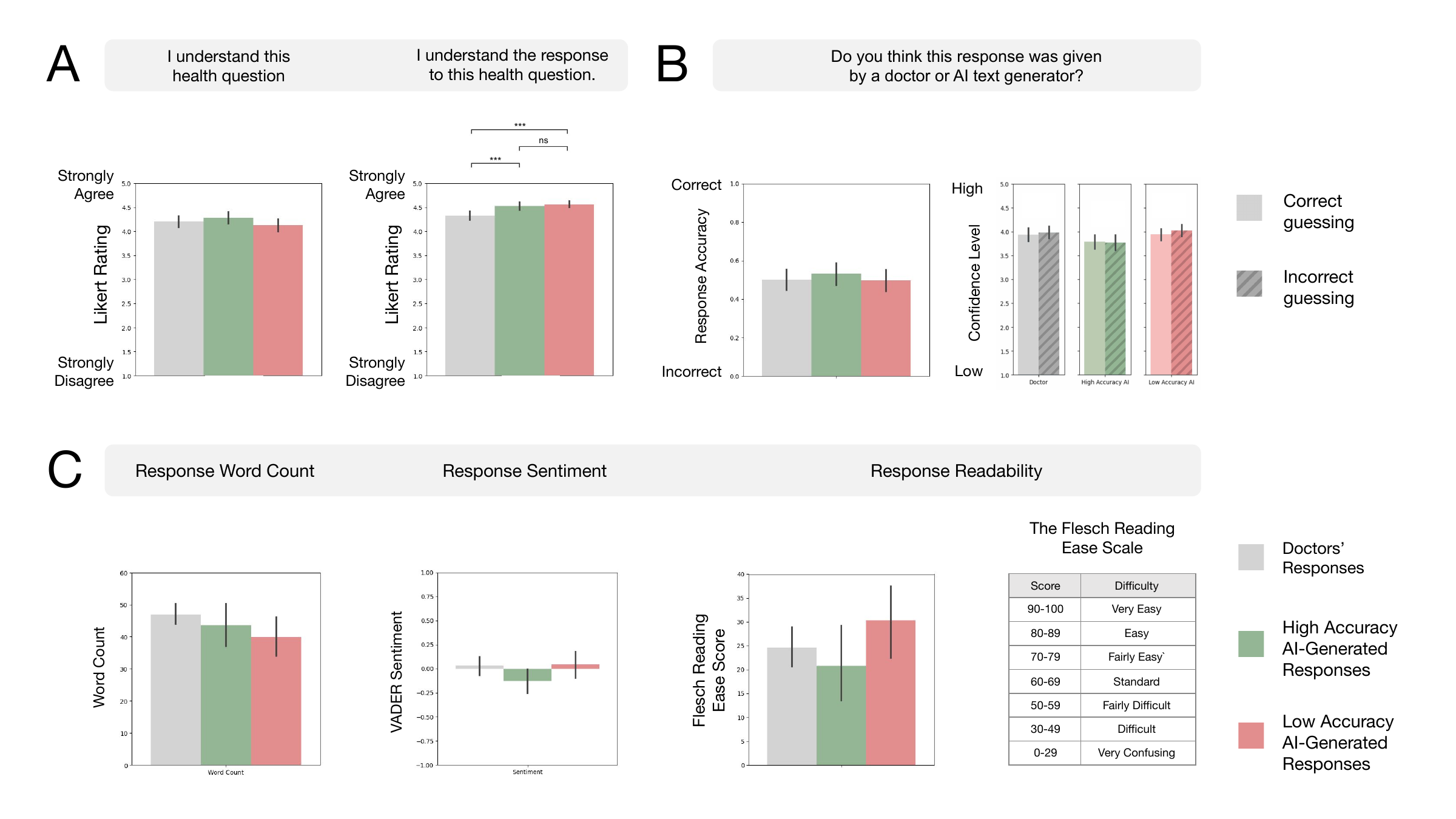}   
    \caption{The participant's ability to distinguish between AI-generated and Doctor-written medical responses. (A) Likert analysis of participant understandability of each medical question and response. (B) Quantification of responses perceived as AI-generated vs. provided by a Doctor in the different response types (Doctor, High Accuracy AI, Low Accuracy AI) with participant confidence levels (C) Analysis of medical response word count, sentiment, and readability.}
    \label{fig:figure-distinguishable}
\end{figure}

\subsubsection*{Understanding Inquiry and Response}
From the hierarchical linear model analysis, there were no significant differences in the participants' understanding of medical questions (p = 0.5964) across the three categories. However, significant differences were observed in participant evaluations of response understanding (p = 1.787e-04). From the pairwise comparison, we found that participants rated their understanding of AI-generated responses, regardless of the accuracy level (High Accuracy AI: mean = 4.53, SD = 0.69;  Low Accuracy AI: mean = 4.56, SD = 0.62), to be significantly higher (High Accuracy AI vs. Doctors: p = 0.001, Low Accuracy AI vs. Doctors: p = 0.0008) than Doctors' responses (mean = 4.32, SD = 0.83).

\subsubsection*{Source Determination Accuracy}
When participants were asked to determine the source of the medical response provided to them (Doctor-written or AI-generated), there was an average source determination accuracy of 50\% for Doctors' responses, 53\% for High AI Accuracy responses, and 50\% for Low AI Accuracy responses. There were no significant differences in the participants' performance in source determination task (p = 0.6539) across the three categories indicating that participants were unable to effectively distinguish Doctors' responses from AI-generated medical responses.

\subsubsection*{Source Determination Confidence}
When participants were asked to provide their level of confidence in determining the source of the medical response provided to them, even though their accuracy was low (around 50\%), participants reported a high level of confidence across three types of response when they answered correctly (Doctors' response: Mean = 3.94, SD = 0.87; High Accuracy AI response: Mean = 3.78, SD = 0.91; Low Accuracy AI response: Mean = 3.94, SD = 0.79) and incorrectly (Doctors' response: Mean = 3.98, SD = 0.84; High Accuracy AI response: Mean = 3.77, SD = 0.93; Low Accuracy AI response: Mean = 4.02, SD = 0.84). The level of confidence when participants guessed correctly and incorrectly was not significantly different across the three response types (Doctors' response: p = 0.6803; High Accuracy AI response: p = 0.9279; Low Accuracy AI response: p = 0.9537). 

\subsubsection*{Linguistic Analysis}
Additionally, a linguistic analysis of the medical responses was completed through a computational approach to identify if there were any significant variations in linguistic characteristics (word count, sentiment. reading ease) in the different response types that could be impacting participant perception. We found that there were no significant differences identified in word count (p = 0.154; Doctors' responses: Mean = 46.9, SD = 20.3; High Accuracy AI Response: Mean = 43.6, SD = 30.1; Low Accuracy AI response: Mean = 40.0, SD = 26.4), VADER sentiment \cite{hutto2014vader} value (p = 0.107; Doctors' responses: Mean = 0.0312, SD = 0.626; High Accuracy AI response: Mean = -0.128, SD = 0.590; Low Accuracy AI response: Mean = 0.0460, SD = 0.559), and Flesch Reading Ease readability score \cite{wrigley2021objective} \cite{szmuda2020readability} (p = 0.250; Doctors' responses: Mean = 24.6, SD = 26.8; High Accuracy AI response: Mean = 20.8, SD = 36.7; Low Accuracy AI Response: Mean = 30.4, SD = 33.2) in the different response types.


\subsection*{Exp.2: Participant's evaluation of AI-generated responses compared to doctors' responses}

Experiment 2 aimed to assess how participants evaluate responses generated by the AI system compared to those provided by doctors when they are unaware of the exact source of the responses. The experiment, similar to Experiment 1, involved 100 participants (96 participants passed the screening and were included in the results)  who were presented with 10 medical question-response pairs randomly selected from a collection of 30 Doctors' responses, 30 High Accuracy AI-generated responses, and 30 Low-Accuracy AI-generated responses.

Here, participants were asked to provide Likert scale evaluations on a scale of 1 (strongly disagree) to 5 (strongly agree) on (1) their understanding of the medical question and (2) their understanding of the response. Additionally, they were asked to indicate their perception of (3) response validity (Yes/No). Finally, participants were asked to provide Likert scale evaluations on (4) the trustworthiness of the response, (5) the completeness and satisfaction of the response, (6) participant tendency to search for additional information based on the response, (7) participant tendency to follow the advice provided in the response, and (8) participant tendency to seek subsequent medical attention as a result of the response.

\subsubsection*{Understanding Inquiry and Response}
From the hierarchical linear model analysis, there were no significant differences in the participants' understanding of medical questions across the three categories: Doctor-written, High Accuracy AI-generated, and Low Accuracy AI-generated (p = 0.43). However, participants demonstrated a significantly higher level of understanding (p = 8.2e-06) of AI-generated responses than the Doctors' responses, regardless of the AI-generated response accuracy level. Participants indicated the highest level of understanding for High Accuracy AI-generated responses (Mean= 4.58, SD=0.73), followed by Low Accuracy AI-generated responses (Mean= 4.48, SD=0.87), and then the Doctors' responses (Mean= 3.97, SD=1.21) (High Accuracy AI vs. Doctor: p < 0.0001; Low Accuracy AI vs. Doctor: p < 0.0001).

\subsubsection*{Validity}
Additionally, significant differences were observed in participant evaluations of response validity within the different response types (p = 0.011). The pairwise analysis indicated that participants perceived the High Accuracy AI-generated (Mean= 0.95, SD=0.22) responses to be significantly more valid (p = 0.0106) than the Doctors’ responses (Mean= 0.81, SD=0.39). The Low Accuracy AI-generated responses  (Mean= 0.87, SD=0.34) performed very comparably to the Doctors’ responses.

\begin{figure}[H]
    \centering
    \includegraphics[width=0.99\textwidth]{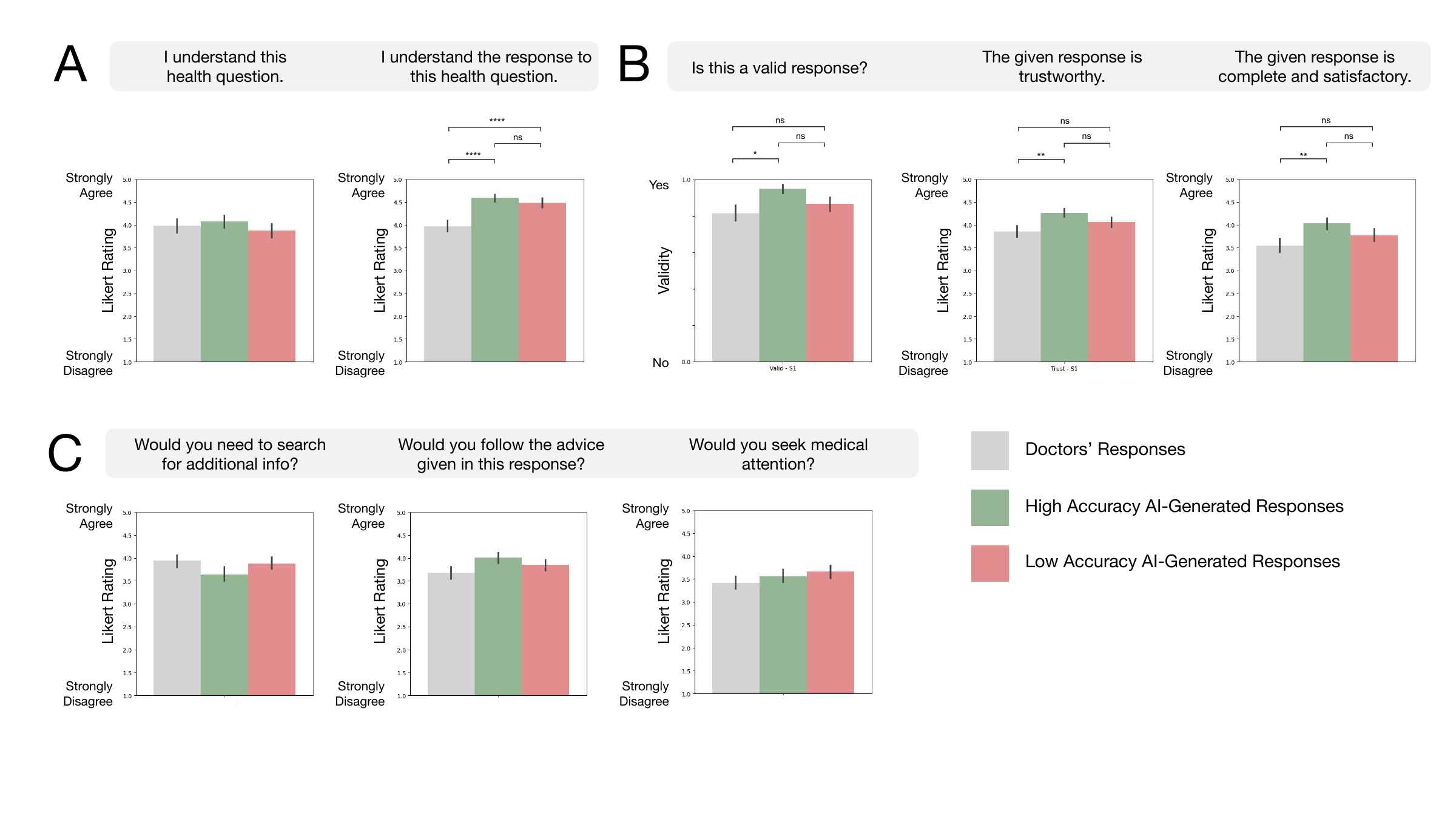} 
    \caption{Participants' perception and evaluation of Doctor-written, High Accuracy AI-generated, and Low Accuracy AI-generated medical responses. (A) Analysis of participant understandability of each medical question and response. (B) Participant evaluation of perceived response validity, trustworthiness, and completeness/satisfaction. (C) Analysis of participant responses to survey questions inquiring about tendency to follow-up, including the likelihood of requiring additional information, following the advice provided in the medical response, and seeking subsequent medical attention.}
    \label{fig:figure-evaluation}
\end{figure}

\subsubsection*{Trustworthiness}
Significant differences were observed in participant evaluations of response trustworthiness within the different response types (p = 0.0058). The pairwise analysis indicated that participants perceived the High Accuracy AI (Mean= 4.26, SD=0.86) to be significantly more trustworthy (p = 0.0050) than the Doctors' responses (Mean= 3.85, SD=1.13). The Low Accuracy AI-generated responses (Mean= 4.06; SD=1.05) were rated similarly to the Doctors’ responses.

\subsubsection*{Completeness/Satisfaction}
Significant differences were observed in participant evaluations of response completeness \&  satisfaction in the different response types (p = 0.005). The pairwise analysis indicated that participants perceived the High Accuracy AI (Mean= 4.03, SD=1.11) to be significantly more complete/satisfactory (p = 0.0042) than the Doctors' responses (Mean= 3.55, SD=1.33). The Low Accuracy AI-generated responses (Mean= 3.77, SD=1.30) were rated similarly to the Doctors’ responses, with no significant difference identified.

\subsubsection*{Tendency to seek additional information}
Beyond the previous metrics, such as validity, trustworthiness, and completeness, we were also interested in gaining an understanding of what next steps the participant might be inclined to take as a result of the response. While these participants were not patients directly seeking the responses to these medical questions, they were asked to emulate the patient in the given situation. Participants told to picture themselves asking their Doctor the given question, were asked to rate their tendency to seek additional information as a result of the response they received. We did not observe significant differences (p = 0.10) between Doctors' responses (Mean= 3.94, SD=1.17), High Accuracy AI response (Mean= 3.65, SD=1.35), and Low Accuracy AI responses (Mean= 3.88, SD=1.21).

\subsubsection*{Tendency to follow the advice provided}
Again, asked to envision themselves as the patient seeking advice, participants rated their tendency to follow the advice provided to them in the response. We did not observe significant differences (p = 0.094) between Doctors' responses  (Mean = 3.68, SD = 1.20), High Accuracy AI response (Mean = 4.00 , SD = 1.02), and Low Accuracy AI responses (Mean = 3.85, SD = 1.14), demonstrating a relatively equal tendency to follow the advice provided across all three response types. 

\subsubsection*{Tendency to seek further medical attention}
Finally, participants were asked to rate their tendency to seek subsequent medical attention as a result of the response provided. We did not observe significant differences (p = 0.26) between Doctors' responses  (Mean = 3.42, SD = 1.31), High Accuracy AI response (Mean = 3.56 , SD = 1.24), and Low Accuracy AI responses (Mean = 3.66, SD = 1.28)

\subsection*{Exp.3: Participant’s evaluation of AI-generated responses compared to doctors’ responses given a random label}

In the third experiment, we investigated if participants exhibited biases toward or against certain response sources. Similar to Experiment 2, 100 participants (all 100 participants passed the screening and were included in the result)  were presented with 10 medical question-response pairs randomly selected from a collection of 30 doctors' responses, 30 High Accuracy AI-generated responses, and 30 Low-Accuracy AI-generated responses. However, at the start of the survey, participants were randomly shown one of three labels: "The responses to each medical question were given by a \%(Doctor)", "The responses to each medical question were given by \%(Artificial Intelligence (A.I.))", or "The responses to each medical question were given by a \%(Doctor assisted by A.I.)".

Then, similar to Experiment 2, participants were asked to provide Likert scale evaluations on a scale of 1 (strongly disagree) to 5 (strongly agree) on (1) their understanding of the medical question and (2) their understanding of the response. They were asked to indicate their perception of (3) response validity (Yes/No). Finally, they were asked to provide Likert scale evaluations on a scale of 1 (strongly disagree) to 5 (strongly agree) of (4) the trustworthiness of the response, (5) the completeness and satisfaction of the response, (6) participant tendency to search for additional information based on the response, (7) participant tendency to follow the advice provided in the response, and (8) participant tendency to seek subsequent medical attention as a result of the response.

The results revealed that, in general, the source labels had little effect on participants' evaluations of the medical responses. However, we observed the effect of the labels on the "trustworthiness" rating of the Doctor's responses (p = 0.022) and High Accuracy AI responses (p = 0.0042), see figure~\ref{fig:figure-lable-influence}. In particular, the pairwise analysis revealed that in the presence of the label: "This response to each medical question was given by a \%(Doctor)" participants tended to rate High Accuracy AI-genered responses types as significantly more trustworthy ("Doctor" vs. "AI": p = 0.013; "Doctor" vs. "Doctor assisted by AI": p = 0.01). However, we didn't see such an impact of the same label ("Doctor") on the trustworthiness ratings of the Low Accuracy AI-generated response (p = 0.49).

\subsection*{Additional Experiment: Physicians' evaluations of AI-Generated responses with and without response source indicated}

\begin{figure}[H]
    \centering
    \includegraphics[width=0.99\textwidth]{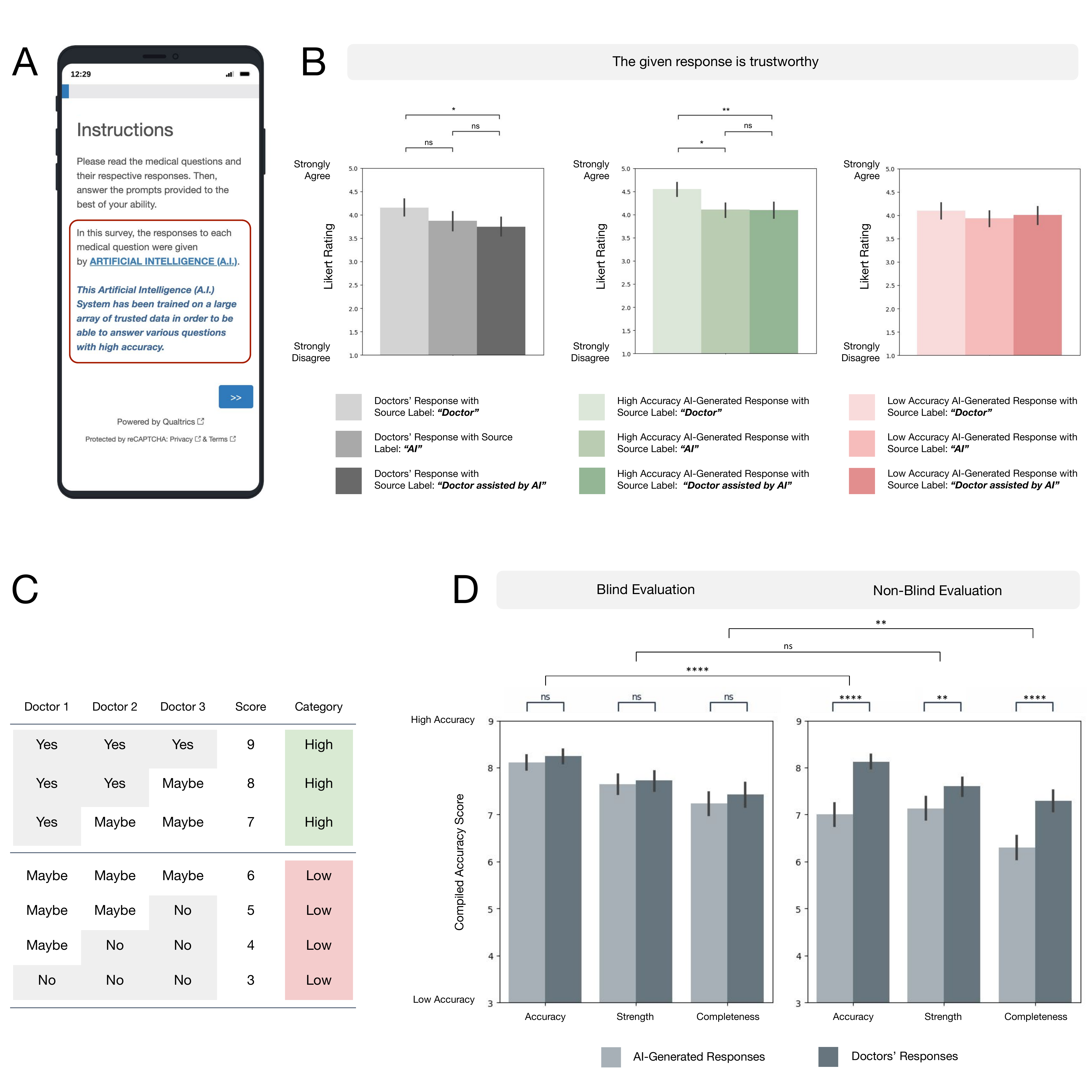}   
    \caption{(A) A screenshot of the survey platform displaying the randomly assigned source label (B) Likert analyses of participant evaluations of response trustworthiness across the different response types and different source labels (“Doctor”, “AI”, “Doctor assisted by AI”). The impact of randomized source labels on participants' perception of Doctor-written, High Accuracy AI-generated, and Low Accuracy AI-generated medical responses. (C) and (D) are part of the Blind and Non-Blind Evaluation of the AI-generated medical responses by physicians. (C) The table represents the compilation of three expert evaluation scores with the values for each evaluation as follows: Yes = 3, Maybe = 2, No = 1. With the following numerical values for each expert evaluation, a compiled score was formed. Any score equal or above 7 (with two or fewer “Maybe” evaluations) is considered High Accuracy. Any score equal to or below 6 (the majority of evaluations are “Maybe” or worse) is considered Low Accuracy. Total of three expert evaluations in each evaluation round, Blind and Non-Blind. (D) Blind Evaluation: Average of 150 compiled scores from each test (Accuracy, Strength, and Completeness) across both response types (AI-Text Generated and Doctor) and Non-Blind Evaluation: Average of 150 compiled scores from each test (Accuracy, Strength, and Completeness) across both response types (AI-Text Generated and Doctor)}
    \label{fig:figure-lable-influence}
\end{figure}

Identifying key results across non-expert participant evaluations of the AI-generated responses vs. the Doctor' responses, we wanted to conduct a preliminary investigation of whether similar trends would be found amongst the physician evaluators. Particularly, we were interested in exploring if our physicians revealed any particular biases during their evaluation of the AI-generated responses. To do so, firstly we asked 3 of our 4 physicians from the initial evaluation to also evaluate the Doctors' responses, completing the Non-Blind portion of the study. Simultaneously, 6 additional general physicians from the same institutions were asked to complete a blind evaluation of the same AI-generated responses and Doctors' responses. More details regarding the design of this additional study can be found in the methodology. We found that when the experts didn't have access to the label regarding the source of the response (Doctor-written or AI-generated), there was no significant difference in their evaluation in terms of accuracy (p = 0.2258), strength (p = 0.5694), and completeness (p = 0.2740). However, when the experts did have access to the source of the response, they evaluated the AI-generated responses as significantly lower in all three metrics: accuracy (p = 6.509e-13), strength (p = 0.003), and completeness (p = 1.606e-08). see figure ~\ref{fig:figure-lable-influence}. Additionally, when completing a two-way ANOVA test, a significant relationship between the study type (Blind vs. Non-Blind) and the Response Source (AI vs. Doctor) was identified while evaluating the Accuracy (p = 1.385e-07) and Completeness (p = 0.001126), confirming a bias presented by experts against AI-generated responses when the source of the response is indicated ~\ref{fig:figure-lable-influence}.

\section*{Discussion}
Firstly, participants present a very similar understanding of the medical questions across the different groups (High Accuracy, Low Accuracy, and Doctor). Therefore, any differences in their subsequent evaluations of the medical responses in each experiment should be associated with differences in perception of the medical response rather than with differences in understanding of the medical question. There were also no significant differences identified in the linguistic characteristics of the different response types, \revision{thus controlling for any confounding factors related to medical question-response linguistics that could impact evaluation outcomes}. 

\subsection*{The general public is unable to distinguish AI-generated medical responses from Doctors responses}
Participants displayed an approximate 50\% accuracy rate in discerning the origin of the medical responses, making it clear that they struggled to effectively differentiate between medical advice offered by a human doctor and medical responses generated by artificial intelligence. This holds true even when the accuracy of the AI-generated medical response is comparatively low. Thus, participants perceive the AI-generated responses as remarkably similar to those provided by doctors, rendering them unable to accurately differentiate between the advice given by the artificial intelligence and that offered by a registered physician on the online healthcare platform HealthTap.

\subsection*{Low Accuracy AI-generated responses pose a danger for the public as they seem equally valid and are deemed more trustworthy}

In addition to participants' inability to distinguish AI-generated responses from doctors' responses, we found that participants evaluated AI-generated responses almost equally to, if not better than, responses provided by Doctors across all metrics. AI-generated medical responses were found to be as comprehensive, valid, trustworthy, complete \& satisfactory, and persuasive as doctors' responses, with AI-generated responses of High Accuracy performing significantly better in a majority of the metrics. Furthermore, on average, albeit not significantly, Low Accuracy AI-generated responses presented a higher level of performance than the doctors' responses across all the evaluation metrics. 

Participants' inability to differentiate between the quality of AI-generated responses and Doctors' responses, regardless of accuracy, combined with their high evaluation of low accuracy AI responses deemed comparable, if not superior, to Doctors' responses, presents a concerning threat. When unaware of the response's source, participants are willing to trust, be satisfied, and even act upon advice provided in AI-generated responses, similar to how they would respond to advice given by a doctor, even when the AI-generated response includes inaccurate information. This unexpected trust and satisfaction with low accuracy AI-generated responses may lead to unwitting acceptance of harmful or ineffective medical advice and concerns of liability for any resulting adverse patient outcomes \cite{maliha21AIliab}.

\subsection*{Participants place significantly more trust in the High Accuracy AI-generated responses that were labeled to suggest they were given by Doctors}

In this study, we found when participants evaluate the medical response provided to them, with no indication of the source, they largely favor AI-generated responses and are ready to place a high level of trust even in those of low accuracy. However, as soon as the source of the response is provided, the way in which participants evaluate the response seems to change. Responses that participants are told were given by Doctors, but were in actuality High Accuracy AI-generated responses were deemed to be more trustworthy than the same High Accuracy AI-generated responses labeled accurately as "AI". In this scenario, we see that the average participant, while satisfied with AI's performance in generating medical advice, seems to \revision{generally} prefer receiving said advice from doctors, and thus reports that medical responses provided by a Doctor are more trustworthy than AI-generated ones.  Interestingly, the presence of the response source label "Doctor" alone does not yield a similar enhancement in the perception of the Low Accuracy AI-generated medical responses. This phenomenon is particularly pronounced when participants are presented with a High Accuracy AI-generated medical response. It is evident that a synergistic effect occurs, wherein the combination of a desirable source with a high accuracy model results in a heightened evaluation. In essence, these two factors need to align to achieve this \revision{desirable} response. This pattern is also observed in other domains, including AI-generated advice for legal decision making. A recent study discovered that while human-like explanations or advice alone did not significantly augment trust, when combined with a high accuracy AI-generated response, there was a notable increase in the level of trust that participants placed in the provided advice \cite{kahr2023seems}. In particular, it is interesting to find this similar bias amongst our expert evaluators, who rated the AI-generated responses significantly higher when the its source was unknown. This finding highlights that even those we rely on to establish the objective truth of our research, assess the efficacy of such models, and endorse their suitability for future use can also be susceptible to their own inherent biases.

\revision{\subsection*{Potential to Extend the Applicability of our Findings to Other Language Models}}
\revision{In our study, we used AI-generated responses from the GPT-3 model, which is among the most adopted language models with publicly accessible specifications and training data \cite{brown2020language}. While there are recent models with greater accuracy, including closed-source models such as GPTs, Claude and Gemini, and open-source alternatives like LLama, we believe that our findings regarding non-experts' perception and evaluation of AI-generated medical responses can be generalized to AI-generated responses from other advanced language models. This is due to the shared underlying architectures and training methodologies across these models that permit newer models to hallucinate misinformation still \cite{nori2023capabilities, bubeck2023sparks, lee2023benefits}. Regardless of the specific language model employed, the possibility of generating both highly accurate and inaccurate medical responses remains a concern \cite{lee2023benefits, ayers2023jama}. As these models progress and refine, the challenges identified in our study, such as lay people's capacity to differentiate between AI-generated and Doctors' medical responses and biases in evaluation, will persist in their relevance and potential even grow. It is critical for future research and development initiatives to take these insights into account when both designing AI models for incorporation into healthcare systems and outlining the framework for their effective, ethical implementation.}

\subsection*{Limitations}
It is important to note that there are some key limitations to this study. (1) As described above, this study uses the GPT-3 model instead of \revision{a more recent version of the model}. An argument could be made that some of the concerns with low accuracy in AI-generated responses may improve with newer models. With that said, it is interesting and concerning to note that even a low accuracy response from an older model was quite convincing to participants. (2) A second limitation to consider is how well do the participants represent the general public. We used an online research recruitment tool which does try to recruit as broad and representative a sample as possible, however this might have skewed towards people who are more technologically savvy and the age range of our participants may have left out older and younger participants as it was mainly participants between 18 - 49 years old. Our participants are not directly affected by the medical advice given, nor are they the ones posing these questions. Instead, they are tasked with emulating the role of a patient who would receive such a response. As a result, while this approach provides a broad understanding of public perception, it lacks the personal element that stems from participants' pre-existing knowledge about the condition and/or their emotional investment in the response provided. (3) Another limitation is that we only had four physicians review each medical question-response pair and evaluate their accuracy. While it would've been ideal to have more expert evaluations, since their scores were ultimately reduced down to two accuracy categories (High and Low), we don't believe this limitation is too critical. (4) A final, crucial, limitation is that this study focuses solely on single medical question and response pairs. It does not deal with conversations or any additional context than what is provided in the inquiry itself. In real world clinical scenarios, Doctors are more likely to ask for additional context and follow up information prior to providing such medical advice. Therefore, further research could explore the role of such context in the application of AI for medical question answering.

\subsection*{Broader Implications}
Our findings expose a number of key considerations that need to be consistently evaluated, both from the perspective of the layperson and the perspective of the physician, while designing and deploying technologies such as LLMs and chatbots in medical response applications.

\subsubsection*{There is a danger in generating and releasing AI-generated medical responses to the public without doctor supervision}
Recent studies regarding the use of LLMs in the medical domain have primarily considered the perspective of the physician or doctors and are limited in assessing the accuracy or quality of the responses \cite{ayers2023jama}. However, our study clarifies that solely identifying the accuracy level of a medical response is inadequate for comprehending the dynamics it may evoke among both general participants and physicians. In the scenario where an AI system predominantly generates medical responses that physicians deem as low accuracy, these experts are likely to label the respective LLM or its design as ineffective for this specific use case. However, using the same LLM and set of medical responses, a lay person may find the advice within those responses highly persuasive and satisfactory, as evidenced in our study, leading them to potentially act upon the advice provided. Consequently, a low accuracy AI-generated response can inadvertently be perceived as favorable or on par with that of a doctor, putting the participant at risk of being harmed by misinformation. This follows a similar pattern to the recent study in which patients increasingly trusted their friends and families over physicians for health issues \cite{Edelman}. Further, our results further support a finding that people cannot detect AI-generated text from those of humans, as the systems can exploit heuristics to produce text perceived as “more human than human”\cite{jakesch2023human}. In cases where such technology is utilized, in order to help reduce the likelihood of such risk, at the minimum a physician should remain within the loop to filter out inaccurate medical information or inappropriate medical advice, while preserving the benefits of the collaboration with the AI system. 

\subsubsection*{Even with a doctor's supervision, it is important to pay attention to how the doctor evaluates the AI system as they could exhibit bias towards or against the AI output}
Both laypersons and experts possess some level of bias with regards to the different medical response types. While lay people find AI-generated responses of high accuracy with a label of "Doctor" to be more trustworthy, expert evaluators find themselves to be more critical of responses labeled as "AI-generated", thus providing another set of considerations that need to be addressed when deploying such systems. The distribution of AI-generated medical information and the level of transparency regarding its source will play crucial roles in evaluating various medical responses. The assessment of the use of AI in the medical domain is far more nuanced than just the one-dimensional evaluation of accuracy. Designing AI-assisted decision-making systems should strike a delicate balance between fostering trust and promoting critical thinking among stakeholders. It is essential to encourage both physicians and lay people not to over-rely on AI, but rather to use AI as an augmentation for engaging in thoughtful evaluation of the provided information. \cite{buccinca2021trust}. 

\subsubsection*{Doctor intervention in Human-AI systems can allow us to benefit from AI's unique capability while preventing the damage from its inaccuracy}
Ultimately, this study shows that participants find the AI-generated responses, especially the High Accuracy responses, to perform comparably if not better than Doctors' responses in all metrics of perception. Thus, lay people tend to very much lean towards AI-generated responses. At the same time participants indicate a clear sense of trust and comfort in responses they are told are given by a Doctor. Thus, ideally we would presume the key for successful applications of LLMs in medicine would be to combine the patient interaction and trust qualities of the physician with the thorough and comprehensive response delivery of the AI. Interestingly enough, we do see though that when participants were told a response was provided by a "Doctor assisted by AI", there was no significance increase in performance evaluations of the response. Thus, the solution of combining the benefits of both physicians and AI seems to be contradicted by participants' perceptions of the different source types. \revision{This underscores the complexities of the situation and emphasizes the intricate dynamics through which participants and experts interact and perceive medical responses}. The multifaceted nature of these perceptions becomes evident, with different details eliciting significantly diverse outcomes, highlighting the nuanced and dynamic aspect of this interaction. Even in a future of "Doctor-assisted by AI" applications, an increase in trust in this particular space would likely rely on the framework or exact method of application. It is crucial to assess all these facets, understand how they are impacted by the given use case, and determine how that influences perception by the public and by experts. This insight would enable us to design future AI systems that promote trust and collaboration among physicians, patients, and AI in time-constrained medical decisions \cite{jacobs2021designing}. Generative AI has found a way to provide information in a very understandable and compelling style. Our results suggest that, when monitored very carefully and effectively, such systems could be used by the medical community to multiply its reach with the lay public while also maintaining the value of an interaction that occurs with a Doctor.

The array of metrics outlined in this paper to assess the interaction between AI and the general participant highlights a number of considerations for practitioners deploying these technologies to keep in mind, including: 1) AI can be promising and seen as a preferable tool for medical response delivery as long as it is accurate. 2) But when it is inaccurate, it could be misleading and misinforming the public because of it's ability to generate persuasive human-like responses. 3) Thus it is important to explore ways to better evaluate the AI and understand how experts can effectively work with AI to expand its potential while eliminating its risk. Thus, in the future healthcare providers should consider understanding the best practices for integration of AI in healthcare delivery systems, outlining the involvement of the physician in the delivery of the AI-generated information and making that transparent and explainable to the patient, and further developing standardized and comprehensive methods for evaluating the collaboration between doctors and AI across different medical domains and applications.

\section*{Conclusion}
Overall, this research provides a new perspective on the discourse regarding the use of AI-text generation in responding to medical questions. AI-generated medical responses are not only indistinguishable by the average person but they are evaluated equally, if not better, than doctors' responses, regardless of their accuracy level. This can pose a danger to the public if AI-generated responses are relied upon solely, without proper expert oversight. False information, misdiagnosis, and incorrect treatment could be provided, leading to harmful or even deadly consequences for individuals seeking advice. While the average participant in blind evaluations demonstrated a clear preference for AI-generated responses in all metrics: understandability, validity, trust, completeness/satisfaction, and persuasion, they particularly indicated a higher level of trust in responses that they believe are provided by a doctor. Ultimately, this study demonstrates that assessing the incorporation of AI in the delivery of medical information is far more multifaceted than we might have originally believed.

\section*{Acknowledgement}
Statistical support was provided by data science specialist Steven Worthington, at the Institute for Quantitative Social Science, Harvard University.

\bibliography{reference}
\setcounter{figure}{0} 
\newpage
\section*{Supplementary Materials}
\subsection*{Methodology}
\subsubsection*{Overview}
This paper presents three experiments investigating AI-generated medical responses to medical questions and whether the AI-generated responses are comparable to physician responses. Additionally, this study explores the perception of these AI-generated medical responses, both from the perspective of the public and physicians. 

\subsubsection*{Dataset Generation}
150 anonymous medical questions and Doctors’ responses were retrieved from question forum on HealthTap (https://www.healthtap.com/), an online healthcare provider. The inquiry covers 6 domains of medicine, including: (1) Preventative \& Risk Factors, (2) Conditions \& Symptoms, (3) Diagnostics \& Tests, (4) Procedures \& Surgeries, (5) Medication \& Treatments, and (6) Recovery \& Wellness, with equal distribution.

\revision{Using GPT-3, Generative Pre-trained Transformer, one of the most adopted language models with publicly accessible specifications and training data \cite{brown2020language}, AI responses were produced for each of the medical questions. We applied the default settings of GPT-3 without any modification to assess the performance of the baseline model (model: text-davinci-2, temperature: 0.7, maximum length: 256).}

\revision{These AI-generated responses were subsequently evaluated by 4 evaluators to establish the ground truth on whether the AI-generated responses were accurate. The evaluators were general physicians recruited from Stanford University and UCSF who were compensated with a \$100 Amazon gift card for their contribution.} Each expert evaluator was asked to evaluate the responses on three axes: accuracy, completeness, and strength. While we primarily focused on the accuracy ratings based on initial clinical testing it was important to provide these additional dimensions of evaluation to ensure we had the best possible accuracy ratings. For example, an AI generated response to an inquiry about treatment for a condition may provide an accurate treatment however it could leave out other possible treatments. When asked to evaluate an answer if the axes of completeness was not available the reviewer may have a more difficult time saying this answer was accurate as it omitted other details, however by providing these axes of evaluation the reviewer could say that this answer is accurate but incomplete. 

Experts were informed that they would be evaluating AI-Generated responses, thus making this a non-blind evaluation. Based on these evaluations made by 4 practicing physicians, providing the ground truth on whether the AI-generated response is correct (Yes = 3, Maybe = 2, No = 1), \revision{compiled} scores for each medical response in a dataset of a hundred and fifty medical question-response pairs were formed. The medical responses were then organized according to the different accuracy levels. Any response with two or less “Maybe” evaluations was considered High Accuracy. Any inquiry with the majority of evaluations being “Maybe” or worse was considered Low Accuracy. This organized dataset was utilized to conduct an array of studies. We used the experts' ratings to select a subset of medical question-response pairs, forming a new dataset consisting of 30 distinct pairs of medical questions and Doctors' responses, 30 distinct pairs of medical questions and High Accuracy AI-generated responses, and 30 distinct pairs of medical questions and Low Accuracy AI-generated responses.

Further on in the study an additional group of six physicians were asked to evaluate the same a hundred and fifty medical question-response pairs but this time with the source of the response unknown (Blind Evaluation). Given evidence of response source influencing participant evaluation of AI-generated responses, we were interested in exploring if our physicians also revealed any particular biases during blind evaluations of AI-generated responses. As such, three of the experts were asked to evaluate dataset A, a dataset of a hundred and fifty medical question-response pairs with a random selection of 75 inquiries answered by a Doctor and the other 75 by AI. The remaining three experts received dataset B, where the same random selection of 75 inquiries this time were answered by AI and the remaining inquiries were answered by a Doctor. 

Overall, there was a total of three expert evaluations for each of the 150 AI-generated Q-R pairs and 150 Doctors' response pairs in both the Blind and Non-Blind experiments. This amounted to a sample of 150 composite expert evaluation scores for AI-generated responses and 150 for Doctors' responses in each of the evaluations (Blind vs. Non-Blind).

\subsubsection*{Task Description}
Firstly, we investigated whether participants would be able to distinguish AI-generated responses from Doctors' responses as a preliminary understanding of participant perception of AI and Doctor in responding to health inquiries. Thus, in the first experiment, 100 online participants were presented with 10 medical question-response pairs randomly selected from a collection of 30 Doctors' responses, 30 High Accuracy AI-generated responses, and 30 Low Accuracy AI-generated responses. After reading the provided medical question-response pair, participants were asked to provide Likert scale evaluations on a scale of 1 (strongly disagree) to 5 (strongly agree) on (1) their understanding of the medical question and (2) their understanding of the response. Additionally, they were asked to indicate (3) their belief of the response source (response given by a \textit{Doctor} or an \textit{AI-Text Generator}) and a Likert scale evaluation of (4) their confidence in the source they selected on a scale of  1 (low confidence) to 5 (high confidence). The full set of questionnaires is listed later in the supplementary material. 

For the second experiment, we aim to assess how participants evaluate responses generated by the AI system compared to those provided by Doctors, when they are unaware of the exact source of the responses. The experiment, similar to Experiment 1, involved 100 participants who were presented with 10 medical question-response pairs randomly selected from a collection of doctors' responses, High Accuracy AI-generated responses, and Low-Accuracy AI-generated responses. Here, participants were asked to provide Likert scale evaluations on a scale of 1 (strongly disagree) to 5 (strongly agree) on (1) their understanding of the medical question and (2) their understanding of the response. Additionally, they were asked to indicate their perception of (3) response validity (Yes/No). Finally, participants were asked to provide Likert scale evaluations on (4) the trustworthiness of the response, (5) the completeness and satisfaction of the response, (6) participant tendency to search for additional information based on the response, (7) participant tendency to follow the advice provided in the response, and (8) participant tendency to seek subsequent medical attention as a result of the response.

In the third experiment, we investigated if participants exhibited biases toward or against certain response types. Similar to Experiments 1 and 2, 100 participants were presented with 10 medical question-response pairs randomly selected from a collection of Doctors' responses, High Accuracy AI-generated responses, and Low-Accuracy AI-generated responses. However, at the start of the survey, participants were randomly shown one of three labels: "The responses to each medical question were given by a \%(Doctor)", "The responses to each medical question were given by \%(Artificial Intelligence (A.I.))", or "The responses to each medical question were given by a \%(Doctor assisted by A.I.)". 

Then, similar to Experiment 2, participants were asked to provide Likert scale evaluations on a scale of 1 (strongly disagree) to 5 (strongly agree) on (1) their understanding of the medical question and (2) their understanding of the response. They were asked to indicate their perception of (3) response validity (Yes/No). Finally, they were asked to provide Likert scale evaluations on a scale of 1 (strongly disagree) to 5 (strongly agree) of (4) the trustworthiness of the response, (5) the completeness and satisfaction  of the response, (6) participant tendency to search for additional information based on the response, (7) participant tendency to follow the advice provided in the response, and (8) participant tendency to seek subsequent medical attention as a result of the response.  

\subsubsection*{Linguistic Analysis}
Additionally, a linguistic analysis of the medical responses was completed to identify if there were any significant variations in linguistic characteristics in the different response types. Analysis of response (1) word count, (2) sentiment, and (3) readability. We used the "Valence Aware Dictionary and sEntiment Reasoner" (VADER) Sentiment Analysis attuned to sentiments expressed in social media \cite{hutto2014vader}. We used VADER sentiment scores based on vaderSentiment library and Flesch Reading Ease readability score based on readability library. We used descriptive statistics and the statistic of variant (ANOVA) to analyze the results.

\subsubsection*{Participants}
\revision{We recruited the participants from an online participant pool using the website Prolific.} Participants were pre-screened to be fluent in English, and older than 18 years old. The study was set to be balanced between male and female participants. 

To ensure valid results, we excluded participants that did not complete the full study and did not pass the screening questions and attention check. After the exclusions, we had 98 participants for experiment 1, 96 participants for experiment 2, and 100 participants for experiment 3. The full demographic data is further down in the supplementary material.

\subsubsection*{Statistical Analysis}
The 30 medical responses from the three conditions (Doctor, High Accuracy AI, and Low Accuracy AI) each had approximately 8-12 participant evaluations. As a result, each condition had a total of approximately 300 participant evaluations. \revision{For experiments 1, 2, and 3, linear mixed effects models, with crossed random effects for subjects and question-response pairs, were used to account for the fact that the participants rated multiple question-response pairs from different response types in random order. This approach allowed for the analysis of multiple evaluations of different medical responses completed by the same participant. The models were constructed in R (version 4.3.1) using the {lme4} package, with the response scores as the dependent variable. An omnibus test of whether average response scores were the same across the three conditions was conducted using a likelihood ratio test. Pairwise comparisons among conditions were calculated using the {emmeans} package in R, with p-values adjusted for family-wise error using the sequential Bonferroni method. The intra-class correlation (ICC) was calculated using the {psych} package in R, with the adjusted ICC reported. For the additional experiment on the physicians’ evaluations of AI-Generated responses with and without response source shown, a simple t-test was utilized to analyze significance between response sources (AI-generated vs. Doctor) within each of the groups (Accuracy ratings from the Blind Study, Accuracy ratings from the Non-Blind Study, and so on). Additionally, to identify if there were statistically significant differences between scores arising across the different test types (Blind vs. Non-Blind) and different response sources (AI-generated vs. Doctor), a two-way ANOVA test was completed.}

\subsubsection*{Approvals} 
This research was reviewed and approved by the MIT Committee on the Use of Humans as Experimental Subjects, protocol number E-4170.

\subsection*{Data Availability}
The raw data will be available on a GitHub repository: https://github.com/mitmedialab/LLM-MedQA

\subsection*{Code Availability}
The code for data cleaning, analysis, and visualization will be available on the same GitHub repository as the data.

\subsection*{Demographic data}
\begin{figure}[H]
    \centering
    \includegraphics[width=0.99\textwidth]{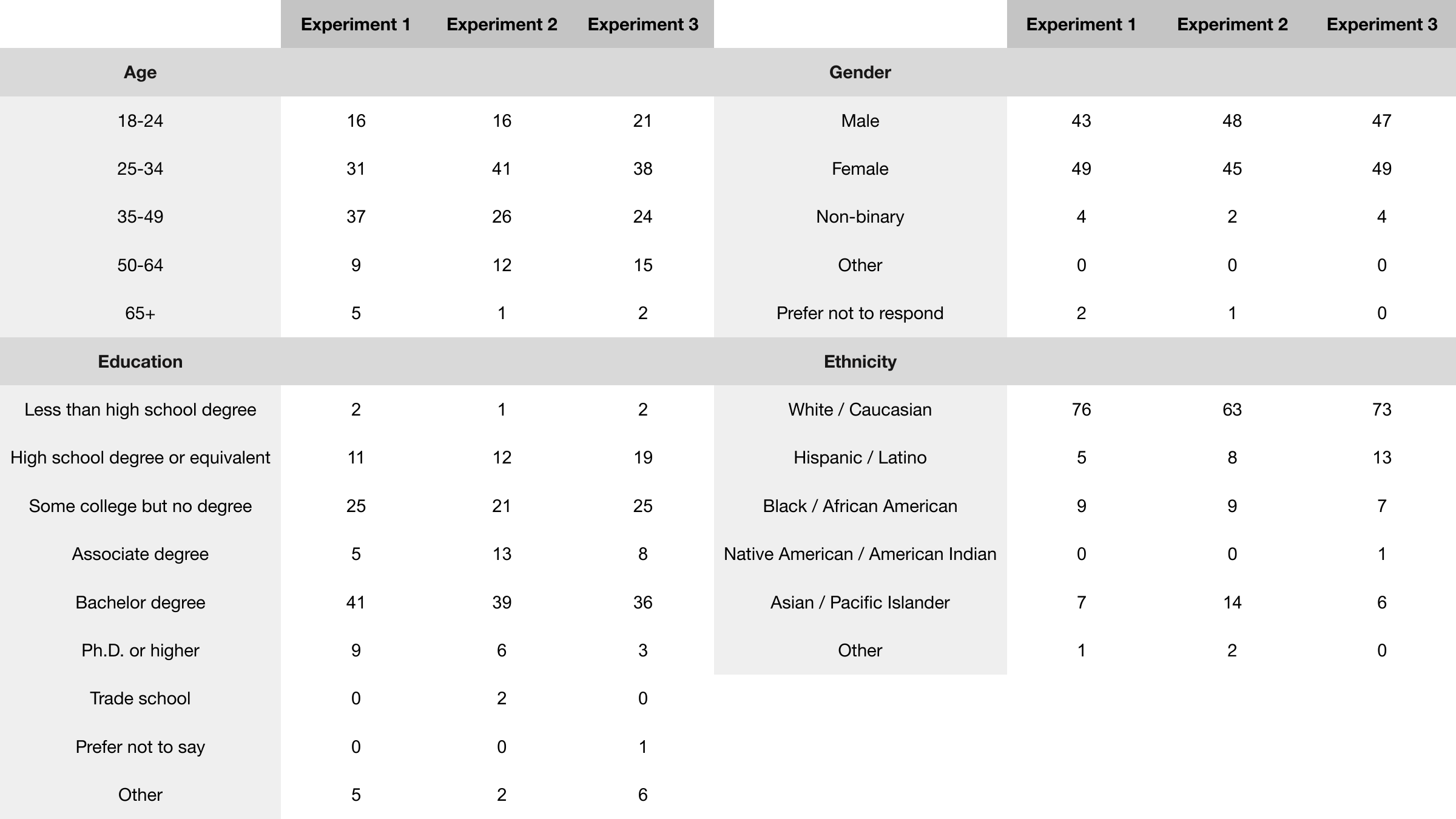} 
    \caption{Demographics of participants in Experiment 1, 2, and 3. Values represent the number of participants in each category. A total of 98 participant evaluations were considered for Experiment 1, 96 for Experiment 2, and 100 for Experiment 3.}
    \label{fig:results}
\end{figure}

\subsection*{Expert Evaluation Questionnaire}

\begin{figure}[H]
    \centering
    \includegraphics[width=0.8\textwidth]{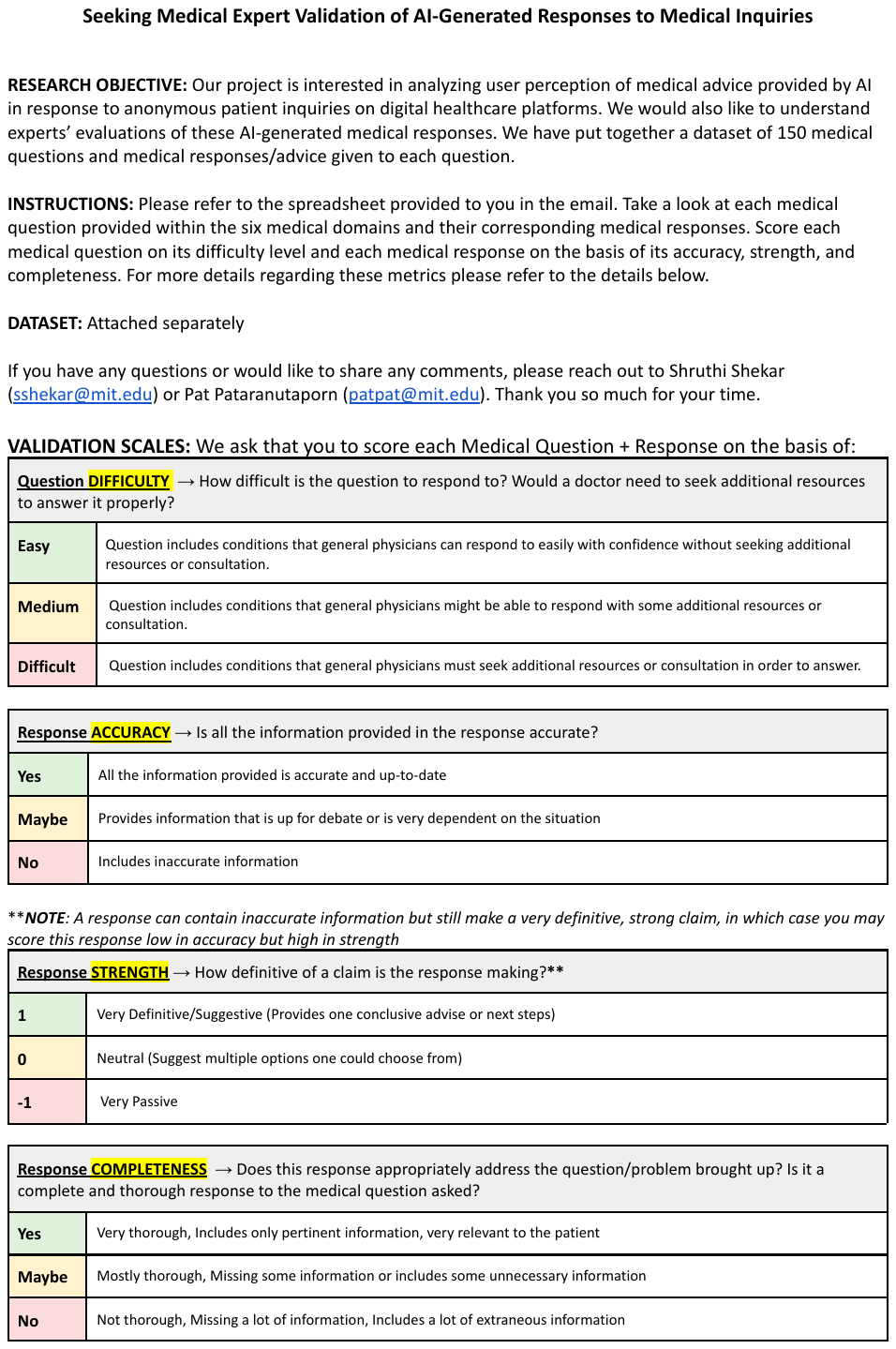}   
    \caption{The expert evaluation's questionnaires and instruction.
}
    \label{fig:expert-questionnaire}
\end{figure}

\newpage

\subsection*{Participant Evaluation Questionnaire}

\subsubsection*{Experiment 1: questions for each question-answer pair}
\begin{itemize}
\item I understand this health question: Likert Scale (Strongly Agree = 5, Strongly Disagree = 1)

\item I understand the response to this health question: Likert Scale (Strongly Agree = 5, Strongly Disagree = 1)

\item Do you think this response was given by a doctor or an AI text generator: Binary Choices(Correct = 1, Incorrect = 0)

\item I am confident in the answer I selected for the
previous question? (Doctor \& AI): Likert Scale (Strongly Agree = 5, Strongly Disagree = 1)

\end{itemize}

\subsubsection*{Experiment 2: questions for each question-answer pair}

\begin{itemize}
\item I understand this health question: Likert Scale (Strongly Agree = 5, Strongly Disagree = 1)

\item I understand the response to this health question: Likert Scale (Strongly Agree = 5, Strongly Disagree = 1)

\item Is this a valid response response?:
Binary Choices(Yes = 1, No = 0)

\item The given response is trustworthy:
Likert Scale (Strongly Agree = 5, Strongly Disagree = 1)

\item The given response is complete and satisfactory: Likert Scale (Strongly Agree = 5, Strongly Disagree = 1)

\item \emph{Imagine you are asking your doctor this question.} After receiving this response, would you need to search for additional info?: 
Likert Scale (Strongly Agree = 5, Strongly Disagree = 1)

\item \emph{Imagine you are asking your doctor this question.} After receiving this response, would you follow the advice given in this response?: 
Likert Scale (Strongly Agree = 5, Strongly Disagree = 1)

\item \emph{Imagine you are asking your doctor this question.} After receiving this response, would you seek medical attention?: 
Likert Scale (Strongly Agree = 5, Strongly Disagree = 1)

\end{itemize}

\subsubsection*{Experiment 3: questions for each question-answer pair}

\begin{itemize}
\item I understand this question: Likert Scale (Strongly Agree = 5, Strongly Disagree = 1)

\item I understand the response to this question: Likert Scale (Strongly Agree = 5, Strongly Disagree = 1)

\item Is this a valid
response?: Binary Choices(Yes = 1, No = 0)

\item The given response
is trustworthy: Likert Scale (Strongly Agree = 5, Strongly Disagree = 1)

\item The given response
is complete and
satisfactory: Likert Scale (Strongly Agree = 5, Strongly Disagree = 1)

\item \emph{Imagine you are asking your doctor this question.} After receiving this response, would you need to search for additional info?: 
Likert Scale (Strongly Agree = 5, Strongly Disagree = 1)

\item \emph{Imagine you are asking your doctor this question.} After receiving this response, would you follow the
advice given in this response?: 
Likert Scale (Strongly Agree = 5, Strongly Disagree = 1)

\item \emph{Imagine you are asking your doctor this question.} After receiving this response, would you seek
medical attention?: 
Likert Scale (Strongly Agree = 5, Strongly Disagree = 1)

\end{itemize}

\subsection*{Statistical report}

\begin{figure}[H]
    \centering
    \includegraphics[width=0.99\textwidth]{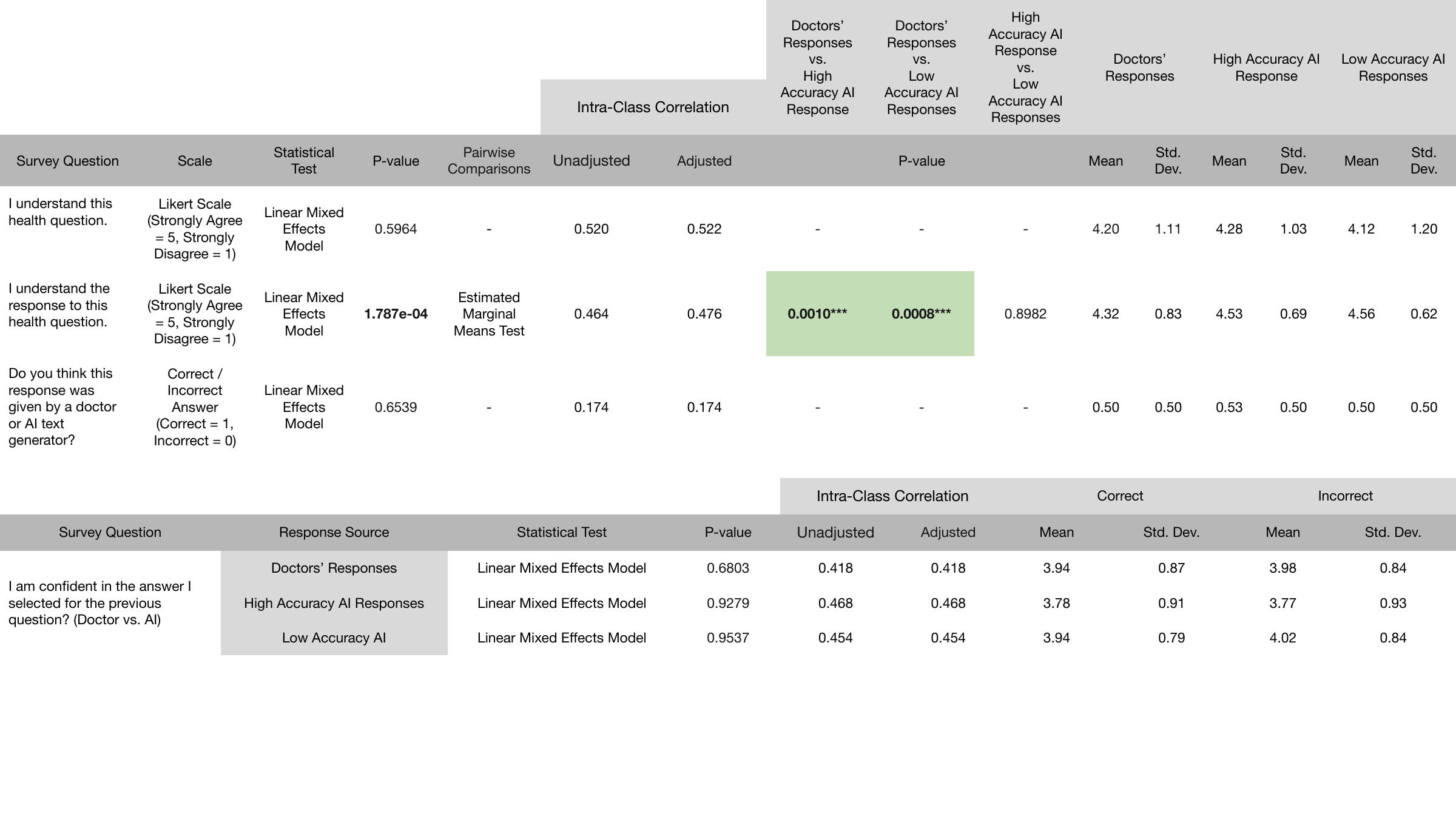} 
    \caption{The full statistical report on participants' ability to distinguish between AI-generated medical responses and Doctors' medical responses from Experiment 1}
    \label{fig:results}
\end{figure}

\begin{figure}[H]
    \centering
    \includegraphics[width=0.99\textwidth]{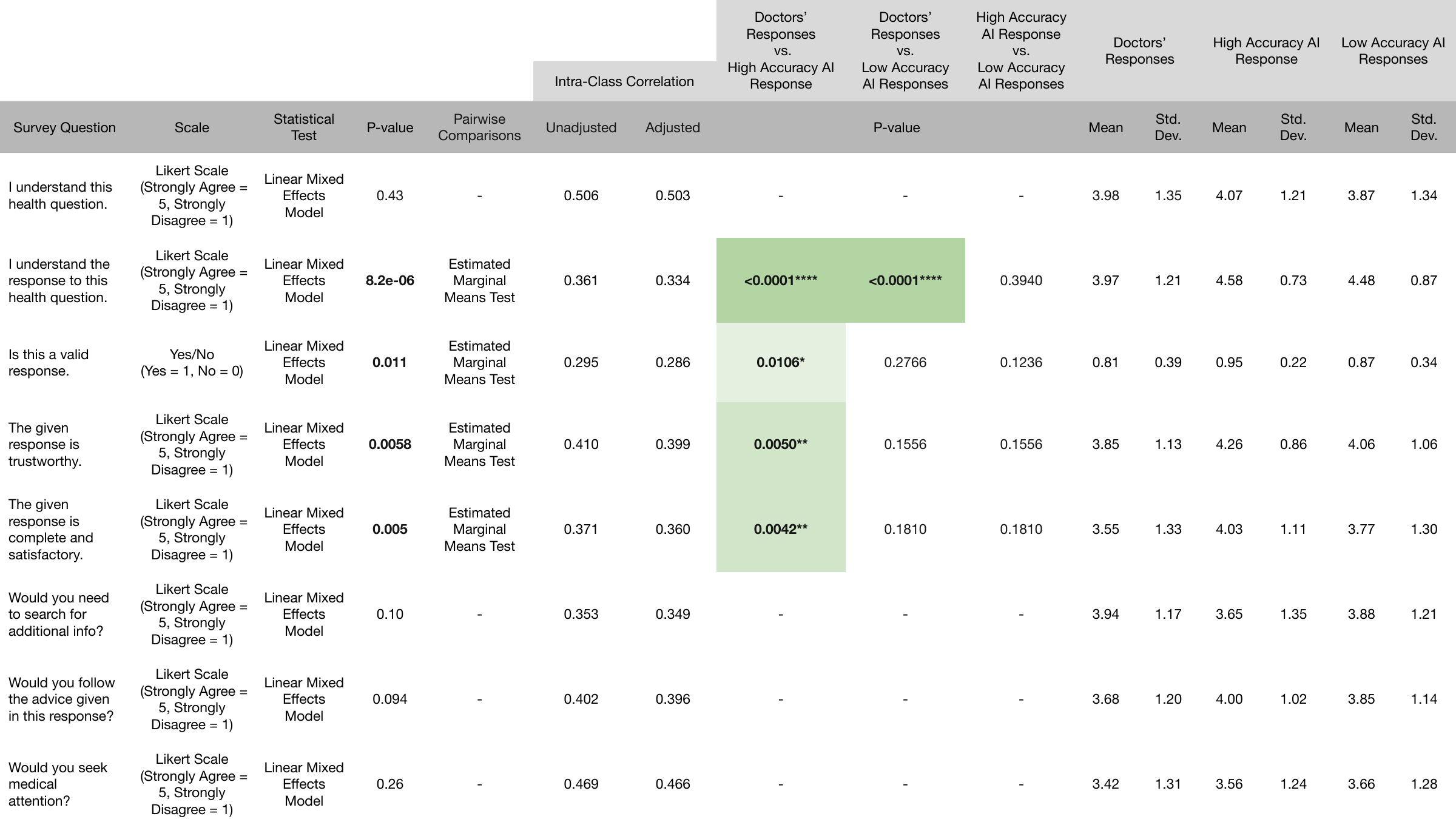} 
    \caption{The full statistical report on participants' perception and evaluation of Doctors' medical responses, High Accuracy AI-generated medical responses, and Low Accuracy AI-generated medical responses from Experiment 2}
    \label{fig:results}
\end{figure}

\begin{figure}[H]
    \centering
    \includegraphics[width=0.99\textwidth]{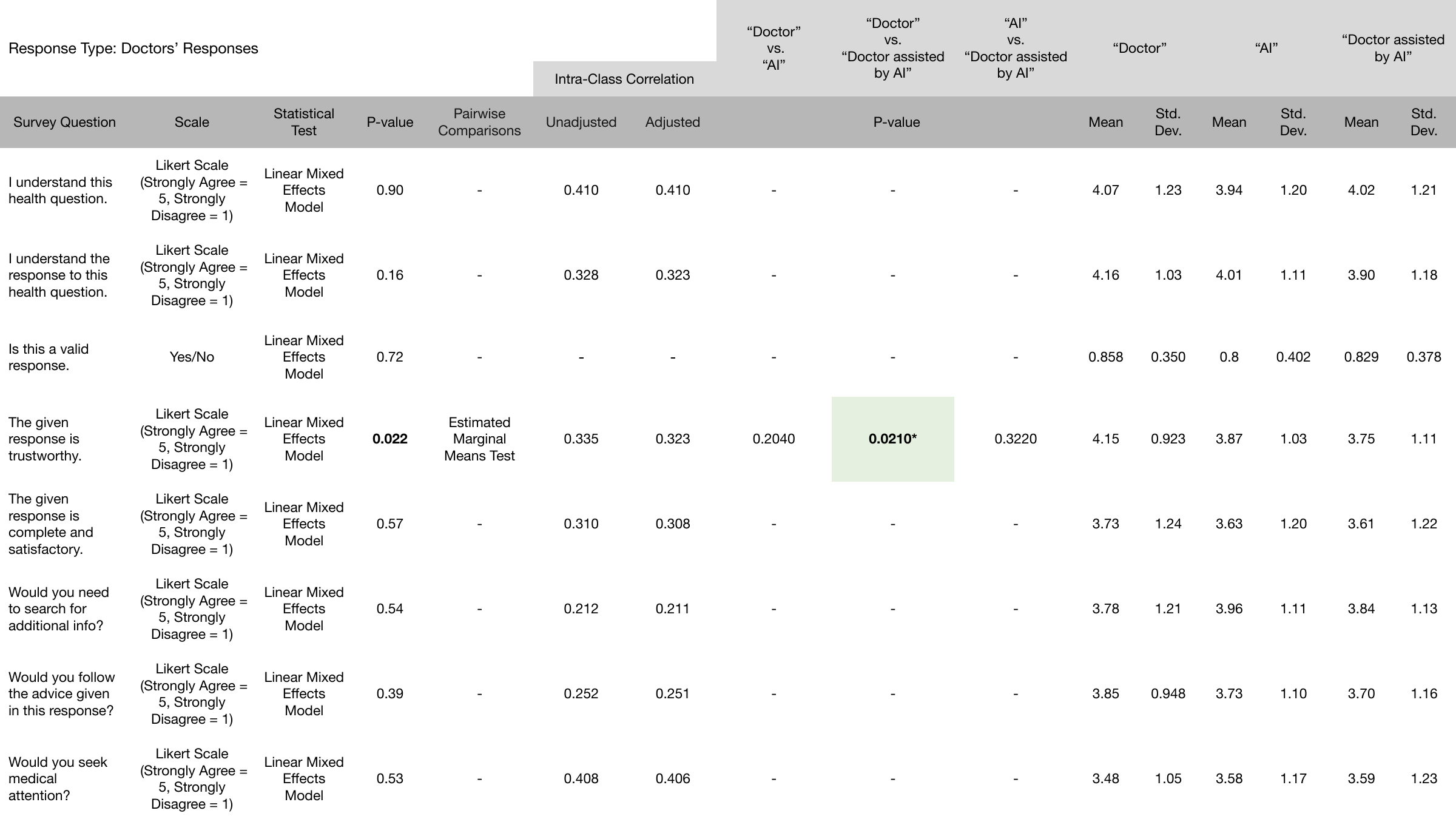} 
    \caption{The impact of the three randomized labels ("Doctor", "AI", "Doctor assisted by AI") on participants' perception of Doctors' medical responses as studied in Experiment 3}
    \label{fig:results}
\end{figure}

\begin{figure}[H]
    \centering
    \includegraphics[width=0.99\textwidth]{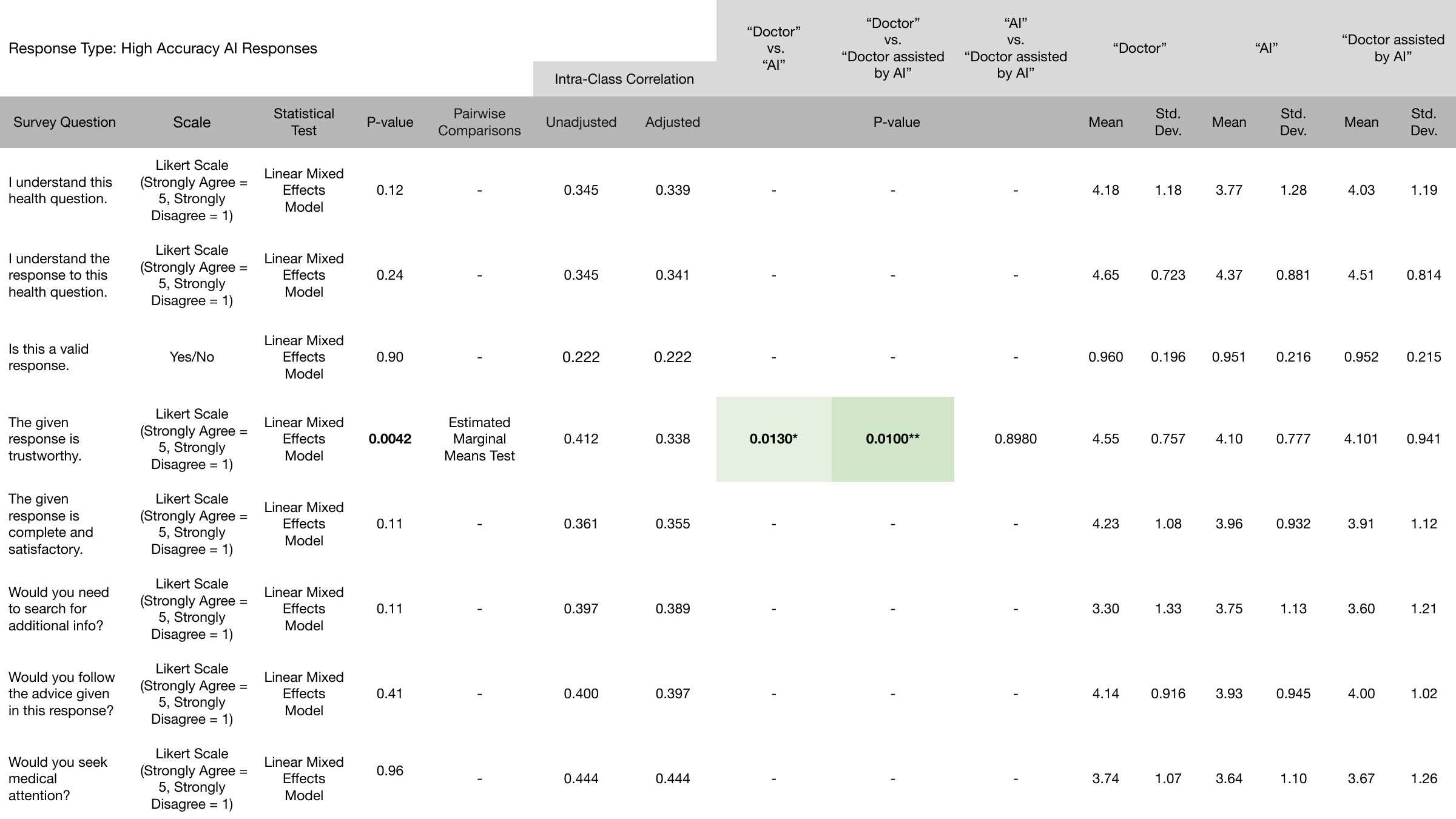} 
    \caption{The impact of the three randomized labels ("Doctor", "AI", "Doctor assisted by AI") on participants' perception of High Accuracy AI-generated medical responses as studied in Experiment 3}
    \label{fig:results}
\end{figure}

\begin{figure}[H]
    \centering
    \includegraphics[width=0.99\textwidth]{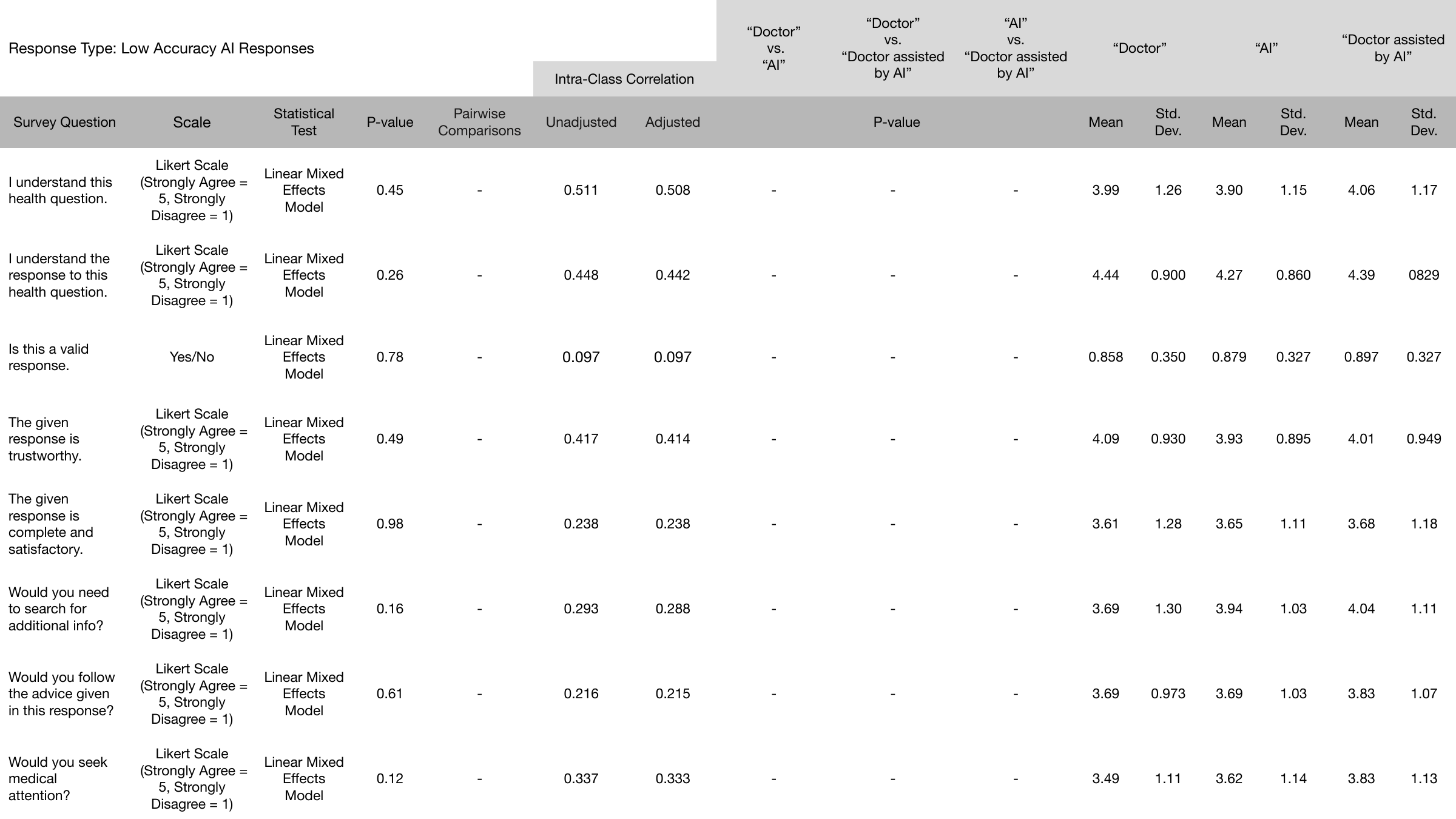} 
    \caption{The impact of the three randomized labels ("Doctor", "AI", "Doctor assisted by AI") on participants' perception of Low Accuracy AI-generated medical responses as studied in Experiment 3}
    \label{fig:results}
\end{figure}

\begin{figure}[H]
    \centering
    \includegraphics[width=0.99\textwidth]{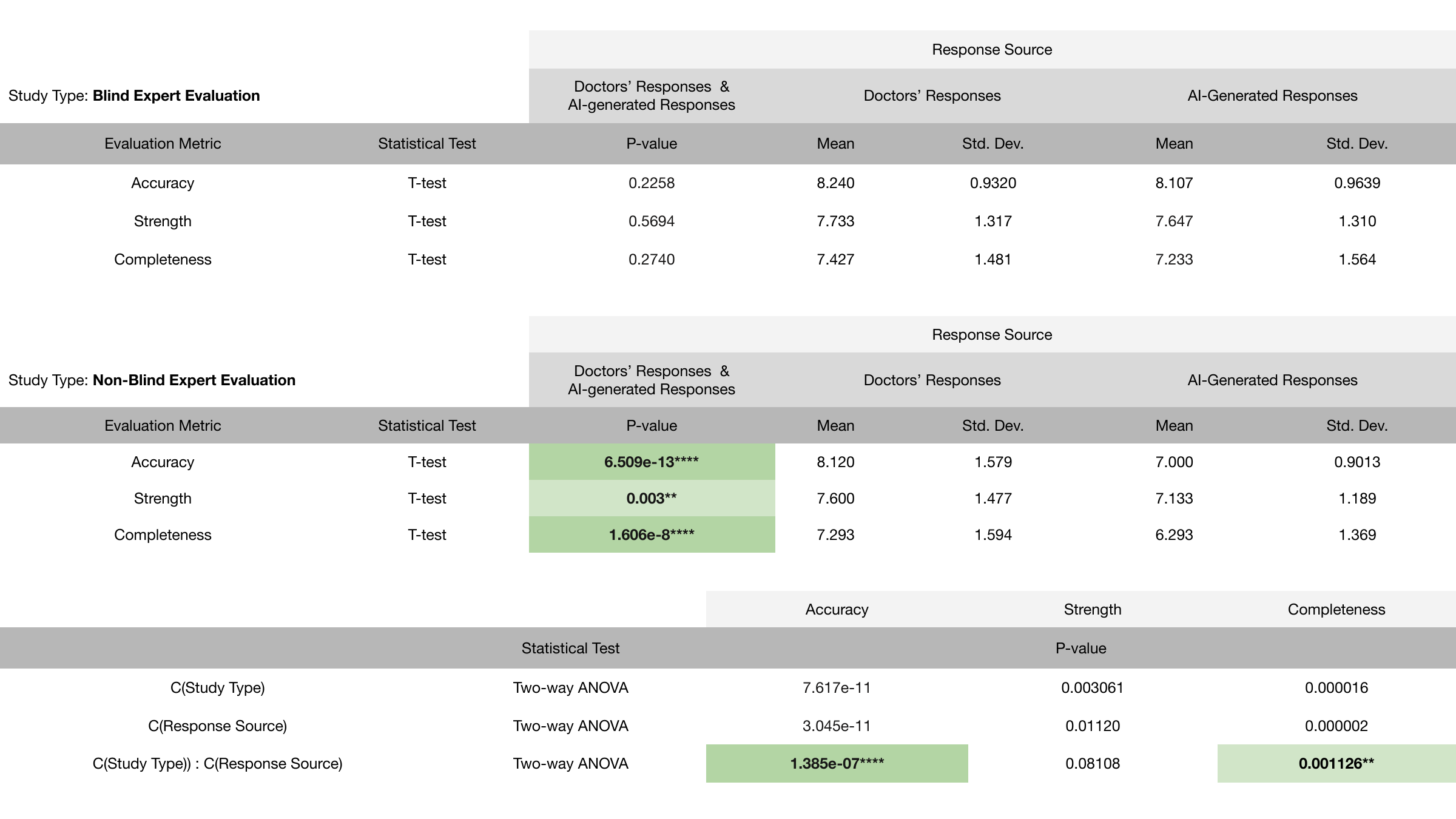} 
    \caption{Expert evaluation of AI-generated response and Doctors' response Accuracy, Strength and Completeness}
    \label{fig:results}
\end{figure}

\begin{figure}[H]
    \centering
    \includegraphics[width=0.99\textwidth]{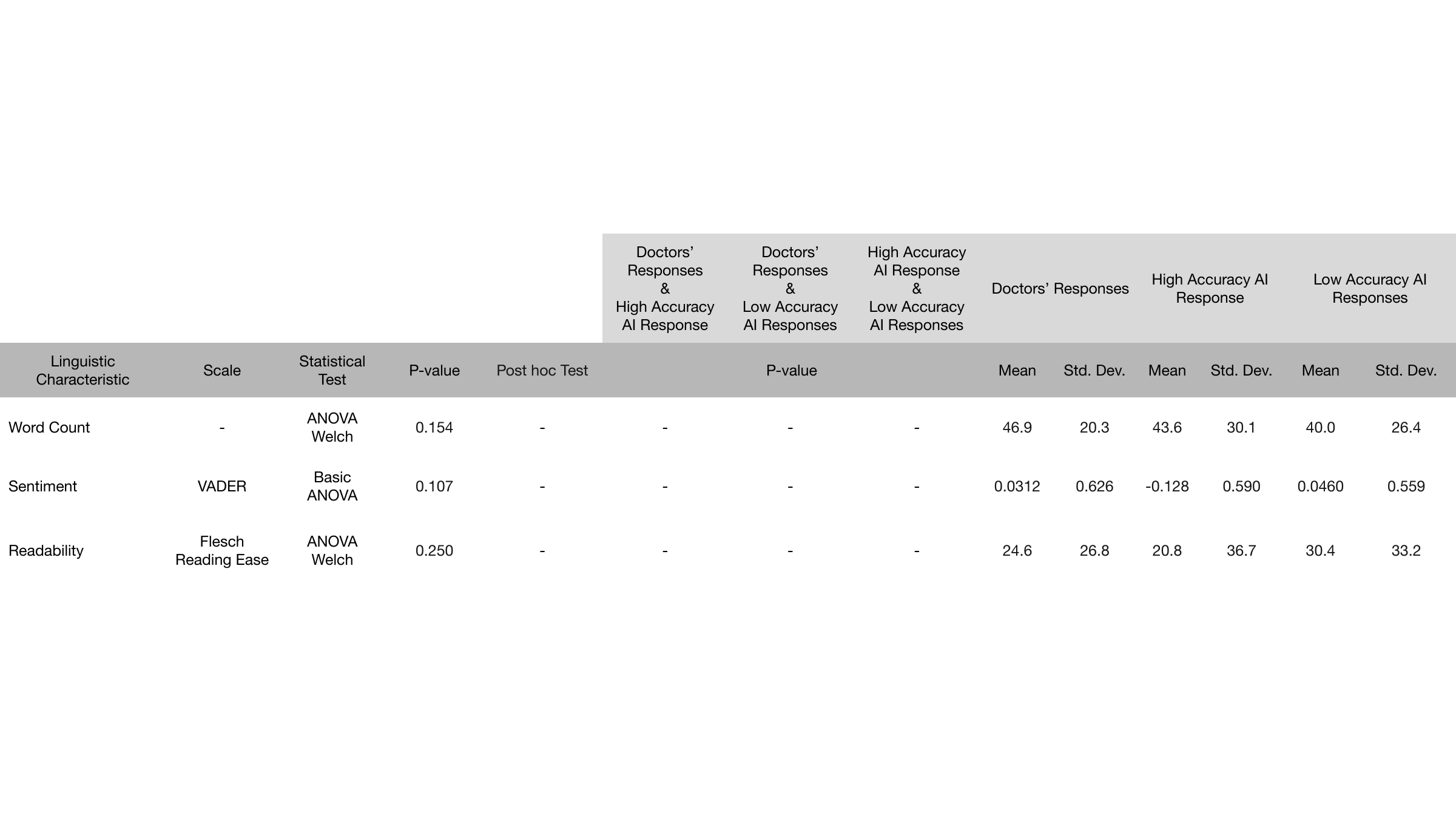} 
    \caption{Linguistic analyses of medical response word count, sentiment, and readability}
    \label{fig:results}
\end{figure}


\end{document}